\RequirePackage{ifpdf}
\ifpdf % We are running pdfTeX in pdf mode
\documentclass[pdftex]{sigma}
\else
\documentclass{sigma}
\fi

\numberwithin{equation}{section}

\usepackage{wick,dsfont}

\begin{document}

\allowdisplaybreaks

\renewcommand{\PaperNumber}{079}

\FirstPageHeading

\ShortArticleName{Non-Perturbative Asymptotic Improvement of Perturbation Theory}

\ArticleName{Non-Perturbative Asymptotic Improvement\\ of Perturbation Theory\\ and Mellin--Barnes Representation}

\Author{Samuel FRIOT~$^{\dag\ddag}$ and David GREYNAT~$^\S$}

\AuthorNameForHeading{S.~Friot and D.~Greynat}

\Address{$^\dag$~Univ Paris-Sud, Institut de Physique Nucl\'eaire,
UMR 8608, Orsay, F-91405, France}
\EmailD{\href{mailto:samuel.friot@ipno.in2p3.fr}{samuel.friot@ipno.in2p3.fr}}
\Address{$^\ddag$~CNRS, Orsay, F-91405, France}

\Address{$^\S$~Institut de F\'isica Altes Energies,
Universitat Aut\`onoma de Barcelona,\\
\hphantom{$^\S$}~E-08193 Bellaterra, Barcelona, Spain}
\EmailD{\href{mailto:greynat@ifae.es}{greynat@ifae.es}}

\ArticleDates{Received June 09, 2010, in f\/inal form September 30, 2010;  Published online October 07, 2010}

\Abstract{Using a method mixing Mellin--Barnes representation and Borel resummation we show how to obtain hyperasymptotic expansions from the (divergent) formal power series which follow from the perturbative evaluation of arbitrary ``$N$-point'' functions for the simple case of zero-dimensional $\phi^4$ f\/ield theory. This hyperasymptotic improvement appears from an iterative procedure, based on inverse factorial expansions, and gives birth to interwoven non-perturbative partial sums whose coef\/f\/icients are related to the perturbative ones by an interesting resurgence phenomenon. It is a non-perturbative improvement in the sense that, for some optimal truncations of the partial sums, the remainder at a given hyperasymptotic level is exponentially suppressed compared to the remainder at the preceding hyperasymptotic level.
The Mellin--Barnes representation allows our results to be automatically valid for a wide range of the phase of the complex coupling constant, in\-cluding Stokes lines.
A~nume\-ri\-cal analysis is performed to emphasize the improved accuracy that this method allows to reach compared to the usual perturbative approach, and the importance of hyperasymptotic optimal truncation schemes.}

\Keywords{exactly and quasi-exactly solvable models; Mellin--Barnes representation; hyperasymptotics; resurgence; non-perturbative ef\/fects; f\/ield theories in lower dimensions}

\Classification{41A60; 30E15}
\vspace{-2mm}

\section{Introduction}

The divergent behavior of a (divergent) asymptotic expansion does not at all detract from its computational utility. This statement is corroborated by the fact that, in what concerns its f\/irst few partial sums, a divergent asymptotic expansion of a given quantity ``converges'' in general much faster to the exact result than what a convergent series representation of the same quantity does. In the case of the Standard Model quantum f\/ield theories, for instance, we may therefore even say that regarding the extreme dif\/f\/iculty to go beyond the f\/irst few perturbative orders when computing observables in QCD or in the electroweak theory, it is an advantage, for phenomenology, to deal with a formalism that leads to presumably\footnote{It is not yet possible to prove that the formal perturbative expansions in the Standard Model quantum f\/ield theories are asymptotic expansions, but the fact that the sum of the f\/irst few perturbative terms is in general in very good agreement with ``exact'' experimental results gives a piece of evidence that perturbation theory is asymptotic to ``something''. The question is: to what? \cite{Fredenhagen:2006rv}} asymptotic power series, diverging for all values of the coupling constants, rather than convergent ones.

\looseness=1
However, one has of course to keep in mind that when dealing with divergent asymptotic perturbative power expansions, there always remains a f\/inite limit of precision beyond which the usual asymptotic theory cannot go, even when the objects that one wants to compute are well-def\/ined\footnote{Still for the case of the (4D) Standard Model of particle physics, due to the absence of a def\/inition of the theory, already the correct evaluation of the size of this precision limit is lacking (i.e.\ theoretical errors implied by truncations of perturbative expansions, OPE, etc. are not under control) although it is in principle of crucial importance in precision test of the Standard Model if one aims to f\/ind new physics ef\/fects.}. All ways to break open this precision limit are welcome. In the beginning of the 1990's,
new asymptotic objects, which have in general a larger region of validity (a larger domain of def\/inition in the complex expansion parameter) and a greater accuracy than conventional asymptotic expansions, appeared in the mathematical literature~\cite{Olver}. With them, a new asymptotic theory emerged: exponential asymptotics (or hyperasymptotics\footnote{The f\/irst paper dealing with hyperasymptotics is~\cite{Berry}, but here we mainly rely on \cite[Chapter~6]{Paris}, where the theory is developed for Mellin--Barnes integrals.}). These asymptotic objects (hyperasymptotic expansions) are very interesting since they correspond, for some optimal truncation schemes to be def\/ined later, to what we could call in physics a~non-perturbative asymptotic improvement of a perturbative (asymptotic) power series. Having in mind future applications in particle or high energy physics, our aim in this paper is to show how hyperasymptotic expansions appear naturally in
the simplest example one may think of, namely zero-dimensional $\phi^4$ f\/ield theory. In particular, we
show that in the course of the study of the $N$-point functions of this theory one may obtain hyperasymptotic expansions directly from the formal (divergent) expansions which follow from their perturbative evaluation and without further information (i.e.\ without using the fact that by their integral representation, for instance, we have a rigorous def\/inition for these objects). It is worth insisting that our f\/inal expressions (hyperasymptotic expansions) are initially obtained following a general and simple non-rigourous approach which is completely justif\/ied at a later stage in the paper. Zero-dimensional $\phi^4$ f\/ield theory has already been used many times to explain new theoretical approaches (see for instance~\mbox{\cite{ZinnJustin:1980uk,Haussling:2001kf}} and, more recently,~\cite{Rivasseau:2009pi}) and we will see that it leads here to non-trivial and interesting issues. We would like to add that although zero-dimensional $\phi^4$ f\/ield theory cannot be considered, strictly speaking, as a realistic toy model in the context of pure particle physics since, for instance, it cannot mimic some of the pathologies of the Standard Model perturbation theory, it is however very likely that the formalism we describe here, by its generality, can be of use in (other f\/ields of) high energy physics (see our conclusions) or in the study of the resummation of higher order corrections in quantum mechanical models and/or superconvergent quantum f\/ield theories that are considered in condensed matter physics (see e.g.~\cite{Nalimov:2009zz}).

The paper is organized as follows. In the introductive Section~\ref{PerturbativeApproach} where basic facts are recalled, the perturbative approach is detailed and, to f\/ix ideas,  numerical results are given for a particular value of the coupling constant $\lambda$. The main part of the paper is Section~\ref{NP}. In Section~\ref{NP1} we present the formal approach which allows to rewrite the perturbative results in terms of hyperasymptotic expansions (at f\/irst level in the hyperasymptotic process). The calculations are based on so-called inverse factorial expansions of the ratios of Euler Gamma functions which constitute the coef\/f\/icients of the perturbative terms forming the tail of the perturbative series. This makes appear the Mellin--Barnes representation into the game. After term by term Borel resummation of the reexpanded tail (this strategy is inspired by \cite[Chapter~21, Section~4]{Dingle} a new expansion of the tail emerges which, added to the perturbative partial sum, form the hyperasymptotic expressions at f\/irst hyperasymptotic level. At this level one may already notice a resurgence phenomenon that links the perturbative coef\/f\/icients with those of the tail's new expansion, and which will also be observed at each higher hyperasymptotic level. Next, Section~\ref{hyperMB} explains in detail how the hyperasymptotic theory of Mellin--Barnes integ\-rals~\cite[Chapter~6]{Paris} leads to the proof that the hyperasymptotic results obtained in Section~\ref{NP1} are correct, and this for a wide range of the phase of the complex coupling constant (taking into account, this time, the integral representation of $N$-point functions that was avoided until here), in particular on Stokes lines. Higher order hyperasymptotic levels are also obtained in this section. In Section~\ref{NP-OTS}, one is concerned with optimal truncation schemes of the hyperasymptotic expansions at zeroth, f\/irst and second hyperasymptotic levels. Their link with the non-perturbative interpretation of our results are underlined. One ends the main body of the paper in Section~\ref{OTS} by performing a numerical analysis which allows to compare the hyperasymptotic expansions for dif\/ferent optimal truncation schemes with the perturbative results. A short appendix give the proof of some results quoted in the text.

\section[Exponential asymptotics in zero-dimensional Euclidean $\phi^4$ theory]{Exponential asymptotics\\ in zero-dimensional Euclidean $\boldsymbol{\phi^4}$ theory}

The 4-dimensional Euclidean $\phi^4$ action is given by
\[
S=\int d^4x\left[\frac{1}{2}\partial_\mu\phi(x)\partial^\mu\phi(x)+\frac{1}{2}m^2\phi^2(x)+\frac{\lambda}{4!}\phi^4(x)\right].
\]
Going to the 0-dimensional theory, which consists to reduce space-time to just one point, makes that the $x$-dependency disappears ($\phi(x)$ becomes a simple real variable $\phi$, the derivative term and the overall integral disappear), so that one gets
\[
S(\phi)=\frac{1}{2}m^2\phi^2+\frac{\lambda}{4!}\phi^4.
\]
The corresponding 4-dimensional generating functional is therefore also reduced, in the 0-dimensional case, to a usual integral of the form
\[
\label{gen}
Z(j)=\mathcal{N}\int_{-\infty}^{+\infty}d\phi\ e^{-\frac{1}{2}m^2\phi^2-\frac{\lambda}{4!}\phi^4+j\phi},
\]
where $j$ is the external source and $\mathcal{N}$ a normalization factor.

In the following, objects under study are the ``$N$-point'' functions, def\/ined as
\begin{gather}
\label{Npoint}
G^{(N)}\doteq\frac{1}{Z(0)}\left.\frac{\partial^NZ(j)}{\partial j^N}\right\vert_{j=0}=\frac{\mathcal{N}}{Z(0)}\int_{-\infty}^{+\infty}d\phi\; \phi^N e^{-\frac{1}{2}m^2\phi^2-\frac{\lambda}{4!}\phi^4},
\end{gather}
where
\begin{gather}
\label{vacuum0}
Z(0)=\mathcal{N}\int_{-\infty}^{+\infty}d\phi\; e^{-\frac{1}{2}m^2\phi^2-\frac{\lambda}{4!}\phi^4}
\end{gather}
is the 0-dimensional version of the generating functional of vacuum to vacuum transitions.

Notice that
\[
G^{(2p+1)} = 0.
\]

Equations~(\ref{Npoint}) (and (\ref{vacuum0})) are def\/ined for ${\rm Re}\,\lambda>0$ and we want to study the small $\lambda$ case. We will see that, thanks to the Mellin--Barnes representation, the results that will be obtained in this paper are in fact valid for a much wider range of complex values of $\lambda$ than just the right half complex $\lambda$-plane. In particular, our results are valid on two Stokes rays (in our case $|\arg \lambda|=\pi$).

Since the theoretical results of interest that we found may already be observed in the study of~$Z(0)$ with $\mathcal{N}=\frac{1}{\sqrt{2\pi}}$, we think that it is more pedagogical to present detailed derivations of our analysis on the simple example of~$Z(0)$ rather than on arbitrary $N$-point functions. This avoids a dependence on the $N$ parameter in the calculations, which would complicate the expressions in the text without really giving new results compared to~$Z(0)$ (apart from the form of the resurgence phenomenon, see Section~\ref{hyperMB}). However, the f\/inal expressions of the calculations will also be given in the more general $N$-point functions case and some subtle changes that have to be done in the computation, for our theoretical strategy to be valid in this more general case, will also be explained when necessary.

\subsection{Perturbative approach}\label{PerturbativeApproach}

In this introductory section, we recall basic facts about the link between the perturbative expansion of 0-dimensional Euclidean $\phi^4$ theory and the counting of Feynman diagrams in the corresponding 4-dimensional theory. We do this on the example of $Z(0)$ and, to f\/ix ideas before the non-perturbative asymptotic analysis, we also give perturbative numerical predictions for a~particular value of the coupling constant.

\subsubsection{Perturbative expansion} \label{perturbative1}

So our f\/irst object of study is $Z(0)$ with\footnote{We could keep the mass as a free parameter but putting it to unity, as well as choosing the normalization constant of Z(0) to be $(2\pi)^{-\frac{1}{2}}$, makes the counting of the 4-dimensional Feynman diagrams, that the 0-dimensional theory allows, more transparent right from the beginning.} $m=1$ and $\mathcal{N}=\frac{1}{\sqrt{2\pi}}$,
\begin{gather}
\label{vacuum}
Z(0)=\frac{1}{\sqrt{2\pi}}\int_{-\infty}^{+\infty}d\phi\ e^{-\frac{1}{2}\phi^2-\frac{\lambda}{4!}\phi^4}.
\end{gather}

In this very simple case, one may get a closed-form expression for (\ref{vacuum}) in terms of special functions or even keep the integral representation and perform the whole (non-perturbative) asymptotic analysis in a fully rigorous way. This will be done as a check, but what we precisely want to show in this paper is a method for performing the analysis in the converse way. Indeed, the question we want to answer is: how can we get a non-perturbative improved expansion from a divergent perturbative expansion when only the latter is available?

Let us compute the perturbative expansion of (\ref{vacuum}) for small $\lambda$ with ${\rm Re}\, \lambda >0$. Replacing $e^{-\frac{\lambda}{4}\phi^4}$ by its series representation and performing term by term integration, one f\/inds
\begin{gather}
\label{asymp}
Z(0)\underset{\lambda\rightarrow 0}{\sim} \frac{1}{\sqrt{\pi}}\sum_{k=0}^{+\infty}\frac{(-1)^k}{k!}\frac{\Gamma\left(\frac{1}{2}+2k\right)}{(3!)^{k}}\lambda^k.
\end{gather}
For further purpose, we def\/ine $u_k$ so that $Z(0)\underset{\lambda\rightarrow 0}{\sim}\sum\limits_{k=0}^{+\infty}u_k$ and the perturbative partial sum $S^{\textrm{Pert}}_{n}\doteq\sum\limits_{k=0}^{n}u_k$.

The f\/irst few terms of (\ref{asymp}) are given by
\begin{gather}
\label{firstfew}
Z(0)=1-\frac{1}{8}\lambda+\frac{35}{384}\lambda^2-\frac{385}{3072}\lambda^3
+\frac{25025}{98304}\lambda^4+\mathcal{O}\big(\lambda^5\big).
\end{gather}

The expansion in the right hand side of (\ref{asymp}) is divergent for \textit{any} value of $\lambda$, as can be seen from the fact that
\begin{gather*}
\left|\frac{u_{k+1}}{u_k}\right|=\left\vert\frac{\frac{(-1)^{k+1}}{(k+1)!}
\frac{\Gamma\left(2(k+1)+\frac{1}{2}\right)}{(3!)^{k+1}}\lambda^{k+1}}{\frac{(-1)^k}{k!}
\frac{\Gamma\left(2k+\frac{1}{2}\right)}{(3!)^{k}}\lambda^k}\right\vert
=\left\vert\frac{1}{k+1}\frac{(2k+\frac{3}{2})(2k+\frac{1}{2})}{3!}\lambda  \right\vert\xrightarrow[k\rightarrow+\infty]{}+\infty,
\end{gather*}
and it is easy to prove that it is an asymptotic expansion of $Z(0)$ as we will see later, but for the moment we only suppose that it is so\footnote{This is why we wrote an asymptotic equality in~(\ref{asymp}).} (we do as in the case of  Standard Model gauge theories where one assumes that perturbative expansions are asymptotic to objects whose def\/inition is still lacking today).

The perturbative result for an arbitrary $N$-point function (with $N=2p$) is similar since
\begin{gather}
\label{intG2p}
\int_{-\infty}^{+\infty}d\phi\; \phi^{2p} e^{-\frac{1}{2}\phi^2-\frac{\lambda}{4!}\phi^4}\underset{\lambda \rightarrow 0}{\sim} 2^{p+\frac{1}{2}}\sum_{k=0}^\infty \frac{(-1)^k}{k!} \; \Gamma\left(\frac{1}{2} + p +2k \right)\left(\frac{\lambda}{3!}\right)^k.
\end{gather}

\subsubsection{Feynman diagrams counting}

\begin{figure}[t]
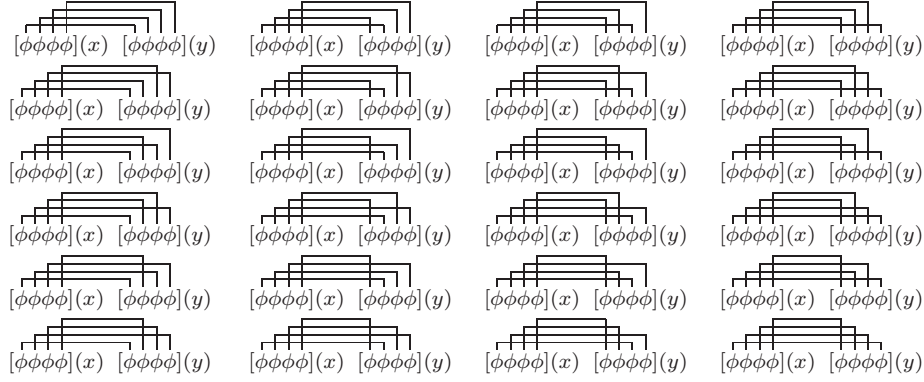

\centering
\begin{tabular}{cccc}
{ \scriptsize$\wick{1234}{[<1\phi<2\phi<3\phi<4\phi](x) \;\;[ >1\phi>2\phi >3\phi >4\phi](y)}$} & {\scriptsize $ \wick{1234}{[<1\phi<2\phi<3\phi<4\phi](x) \;\;[ >2\phi>1\phi >3\phi >4\phi](y)}$} & {\scriptsize $ \wick{1234}{[<1\phi<2\phi<3\phi<4\phi](x) \;\;[ >2\phi>3\phi >1\phi >4\phi](y)}$} & {\scriptsize $ \wick{1234}{[<1\phi<2\phi<3\phi<4\phi](x) \;\;[ >2\phi>3\phi >4\phi >1\phi](y)}$} \\
{\scriptsize $ \wick{1234}{[<1\phi<2\phi<3\phi<4\phi](x) \;\;[ >1\phi>2\phi >4\phi >3\phi](y)}$} & {\scriptsize $ \wick{1234}{[<1\phi<2\phi<3\phi<4\phi](x) \;\;[ >2\phi>1\phi >4\phi >3\phi](y)}$} & {\scriptsize $ \wick{1234}{[<1\phi<2\phi<3\phi<4\phi](x) \;\;[ >2\phi>4\phi >1\phi >3\phi](y)}$} & {\scriptsize $ \wick{1234}{[<1\phi<2\phi<3\phi<4\phi](x) \;\;[ >2\phi>4\phi >3\phi >1\phi](y)}$} \\
{\scriptsize  $\wick{1234}{[<1\phi<2\phi<3\phi<4\phi](x) \;\;[ >1\phi>3\phi >2\phi >4\phi](y)}$} & {\scriptsize $ \wick{1234}{[<1\phi<2\phi<3\phi<4\phi](x) \;\;[ >3\phi>1\phi >2\phi >4\phi](y)}$} & {\scriptsize $ \wick{1234}{[<1\phi<2\phi<3\phi<4\phi](x) \;\;[ >3\phi>2\phi >1\phi >4\phi](y)}$} & {\scriptsize $ \wick{1234}{[<1\phi<2\phi<3\phi<4\phi](x) \;\;[ >3\phi>2\phi >4\phi >1\phi](y)}$} \\
{\scriptsize $ \wick{1234}{[<1\phi<2\phi<3\phi<4\phi](x) \;\;[ >1\phi>3\phi >4\phi >2\phi](y)}$} & {\scriptsize $ \wick{1234}{[<1\phi<2\phi<3\phi<4\phi](x) \;\;[ >3\phi>1\phi >4\phi >2\phi](y)}$} & {\scriptsize $ \wick{1234}{[<1\phi<2\phi<3\phi<4\phi](x) \;\;[ >3\phi>4\phi >1\phi >2\phi](y)}$} & {\scriptsize $ \wick{1234}{[<1\phi<2\phi<3\phi<4\phi](x) \;\;[ >3\phi>4\phi >2\phi >1\phi](y)}$} \\
{\scriptsize $ \wick{1234}{[<1\phi<2\phi<3\phi<4\phi](x) \;\;[ >1\phi>4\phi >2\phi >3\phi](y)}$} & {\scriptsize $ \wick{1234}{[<1\phi<2\phi<3\phi<4\phi](x) \;\;[ >4\phi>1\phi >2\phi >3\phi](y)}$} & {\scriptsize $ \wick{1234}{[<1\phi<2\phi<3\phi<4\phi](x) \;\;[ >4\phi>2\phi >1\phi >3\phi](y)}$} & {\scriptsize $ \wick{1234}{[<1\phi<2\phi<3\phi<4\phi](x) \;\;[ >4\phi>2\phi >3\phi >1\phi](y)}$} \\
{\scriptsize $ \wick{1234}{[<1\phi<2\phi<3\phi<4\phi](x) \;\;[ >1\phi>4\phi >3\phi >2\phi](y)}$} & {\scriptsize $ \wick{1234}{[<1\phi<2\phi<3\phi<4\phi](x) \;\;[ >4\phi>1\phi >3\phi >2\phi](y)}$} & {\scriptsize $ \wick{1234}{[<1\phi<2\phi<3\phi<4\phi](x) \;\;[ >4\phi>3\phi >1\phi >2\phi](y)}$} & {\scriptsize  $ \wick{1234}{[<1\phi<2\phi<3\phi<4\phi](x) \;\;[ >4\phi>3\phi >2\phi >1\phi](y)}$} \\
\end{tabular}
\caption{\label{contractions}The $4!$ dif\/ferent Wick contractions leading to the topology of last diagram in~(\ref{graphelambda2}).}
\end{figure}

It is clear that the coef\/f\/icients in the expansion (\ref{asymp}) are related to the counting of Feynman diagrams in the 4-dimensional theory (taking into account their symmetry factors), since at f\/irst order in the $\lambda$-expansion we have, for 4-dimensional $Z(0)$, the contribution
\begin{gather}
\label{graphelambda1}
\frac{(-1)}{1!}\lambda \frac{3}{4!} \; \parbox{0.8cm}{\includegraphics[width=0.04\textwidth]{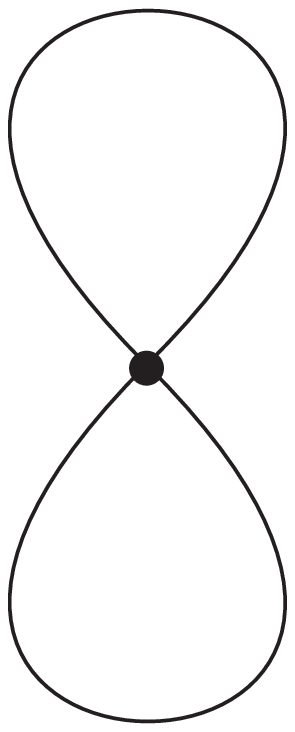}} = - \frac{1}{8}  \lambda \;  \parbox{0.8cm}{\includegraphics[width=0.04\textwidth]{Friot-graphe1}}
\end{gather}
and, at second order,
\begin{gather}
\frac{(-1)^2}{2!}\frac{6\times 6\times 2}{(4!)^2}  \lambda^2   \left( \; \parbox{1.6cm}{\includegraphics[width=0.04\textwidth]{Friot-graphe1} \;\includegraphics[width=0.04\textwidth]{Friot-graphe1} } \right) + \frac{(-1)^2}{2!}\frac{3\times 3}{(4!)^2}   \lambda^2\; \parbox{0.65cm}{\includegraphics[width=0.04\textwidth]{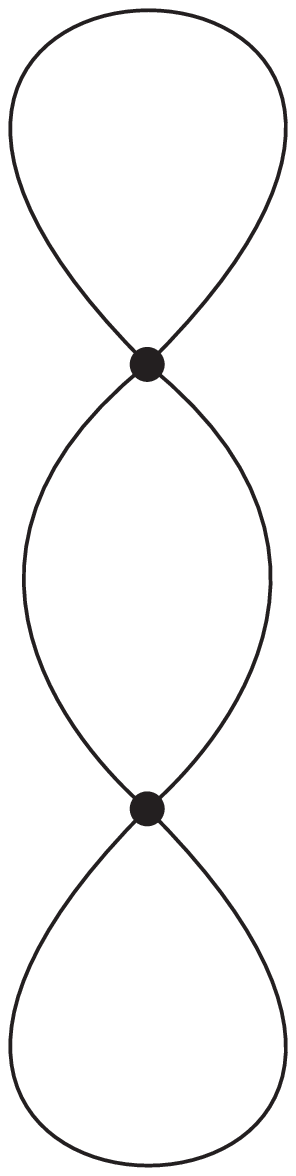}} + \frac{(-1)^2}{2!}\frac{4!}{(4!)^2}  \lambda^2\; \parbox{0.8cm}{\includegraphics[width=0.08\textwidth]{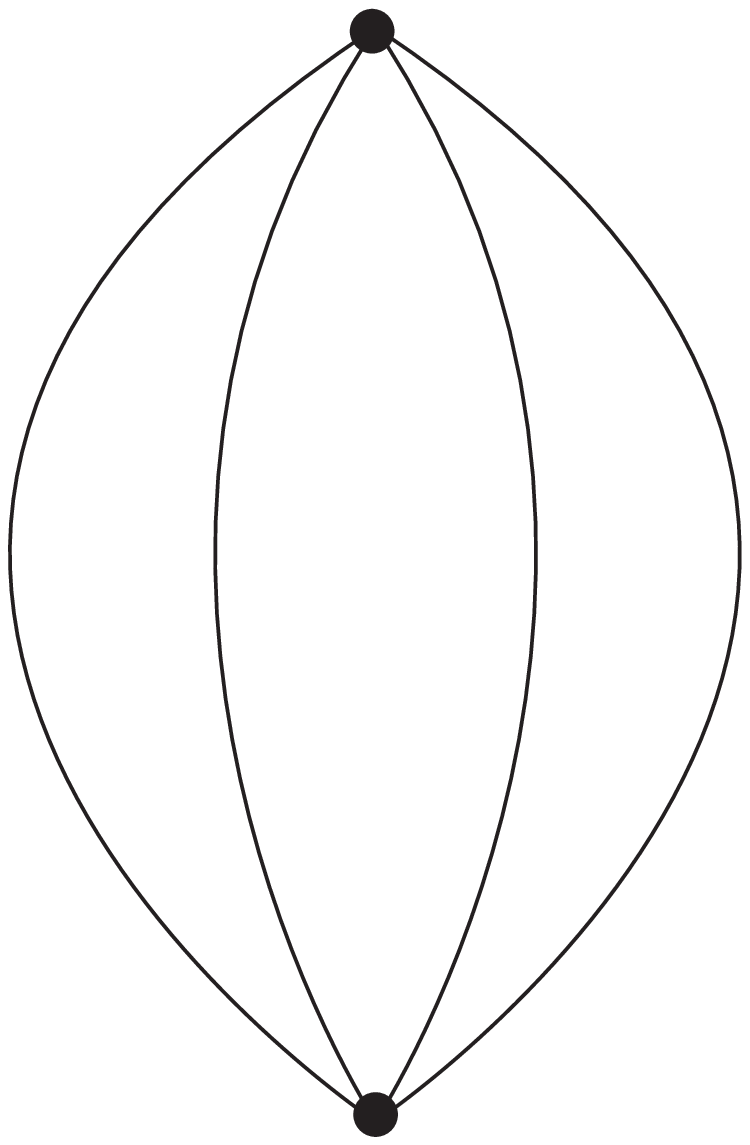}} \nonumber\\
\qquad {}=\frac{24}{384}  \lambda^2   \left( \; \parbox{1.6cm}{\includegraphics[width=0.04\textwidth]{Friot-graphe1} \;\includegraphics[width=0.04\textwidth]{Friot-graphe1} } \right) + \frac{3}{384}   \lambda^2\; \parbox{0.65cm}{\includegraphics[width=0.04\textwidth]{Friot-graphe2}} + \frac{8}{384}  \lambda^2\; \parbox{1.5cm}{\includegraphics[width=0.08\textwidth]{Friot-graphe3}}\label{graphelambda2}
\end{gather}
where, as an example of how we get the coef\/f\/icients in the left hand side of (\ref{graphelambda2}), the 4! in the numerator of the coef\/f\/icient in front of the last diagram is the number of dif\/ferent Wick contractions of the f\/ields that lead to this topology (see Fig.~\ref{contractions}).

We see that by summing the coef\/f\/icients of each dif\/ferent topology (and this def\/ines our counting of Feynman diagrams), one obtains from (\ref{graphelambda1}), (\ref{graphelambda2}) and higher orders the result written in (\ref{asymp}), see also (\ref{firstfew}), apart from the f\/irst term that does not exist in the 4-dimensional theory, since there are no tree vacuum diagrams.

Zero-dimensional f\/ield theories therefore have a practical interest for 4-dimensional particle phenomenology by the fact that they allow a partial but important check of Feynman diagram coef\/f\/icients appearing in perturbative calculations.

\subsubsection{Numerical analysis}\label{pertNum}

Let us now perform some numerical analysis to see the ef\/f\/iciency and predictive power of the perturbation theory developed in the preceding subsections.
In the following, we choose $\lambda=\frac{1}{3}$.

Imagine that we do not know the value of (\ref{vacuum}) for $\lambda=\frac{1}{3}$ so that the only information that we have for computing $Z(0)$ is the divergent perturbative expansion in (\ref{asymp}) (remember that we made the hypothesis that it is an asymptotic expansion of Z(0)). Let us see what is the best perturbative prediction that can be obtained from (\ref{asymp}). In Table \ref{table1}, we computed the f\/irst twenty truncated sums $S^{\textrm{Pert}}_{n-1}$ ($n\in\{1,\dots,20\}$) of (\ref{asymp}) for $\lambda=\frac{1}{3}$ (and the values of the general term of the perturbative series for the same values of $n$).
\begin{table}[h]\centering
\caption{\label{table1}Numerical values of the perturbative general term and partial sums of (\ref{asymp}), for $\lambda=\frac{1}{3}$, with an 8 decimal places precision.}
\vspace{1mm}

\begin{tabular}{|c||c||c||c|}
\hline
$k$ & $u_k$ & $n$ & $S^{\textrm{Pert}}_{n-1} = \sum\limits_{k=0}^{n-1} u_k$\\
\hline
 0 & $\phantom{-}$1.00000000 & 1 & 1.00000000 \\
 1 & $-0.04166667$ & 2 & 0.95833333 \\
 2 & $\phantom{-}$0.01012732 & 3 & 0.96846065 \\
 3 & $-0.00464169$ & 4 & 0.96381896 \\
 4 & $\phantom{-}$0.00314281 & 5 & 0.96696177 \\
 5 & $-0.00281980$ & 6 & 0.96414197 \\
 6 & $\phantom{-}$0.00315269 & 7 & 0.96729466 \\
 7 & $-0.00422235$ & 8 & 0.96307231 \\
 8 & $\phantom{-}$0.00659010 & 9 & 0.96966241 \\
 9 & $-0.01174624$ & 10 & 0.95791617 \\
  \hline
\end{tabular} \hspace*{0.2cm}
\begin{tabular}{|c||@{}c||c||@{}c|}
\hline
$k$ & $u_k$ & $n$ & $S^{\textrm{Pert}}_{n-1} = \sum\limits_{k=0}^{n-1} u_k$\\
\hline
 10 & $\phantom{-0000}$0.02354141 & 11 & $\phantom{-0000}$0.98145758 \\
 11 & $\phantom{0000}{-}0.05240343$ & 12 & $\phantom{-0000}$0.92905416 \\
 12 & $\phantom{-0000}$0.12827922 & 13 & $\phantom{-0000}$1.05733338 \\
 13 & $\phantom{0000}{-}0.34248906$ & 14 & $\phantom{-0000}$0.71484431 \\
 14 & $\phantom{-0000}$0.99043216 & 15 & $\phantom{-0000}$1.70527648 \\
 15 & $\phantom{0000}{-}3.08409571$ & 16 & $\phantom{0000}{-}1.37881923$ \\
 16 & $\phantom{-00}$10.2883505 & 17 & $\phantom{-0000}$8.90953129 \\
 17 & $\phantom{00}{-}36.6060184$ & 18 & $\phantom{00}{-}27.6964871$ \\
 18 & $\phantom{-}$138.374139 & 19 & $\phantom{-}$110.677652 \\
 19 & $-553.800008$ & 20 & $-443.122356$ \\
% 20 & 2339.42045 & 1896.29810 \\
 \hline
\end{tabular}
\end{table}

\begin{figure}[t]
\centering
\includegraphics[width=0.6\textwidth]{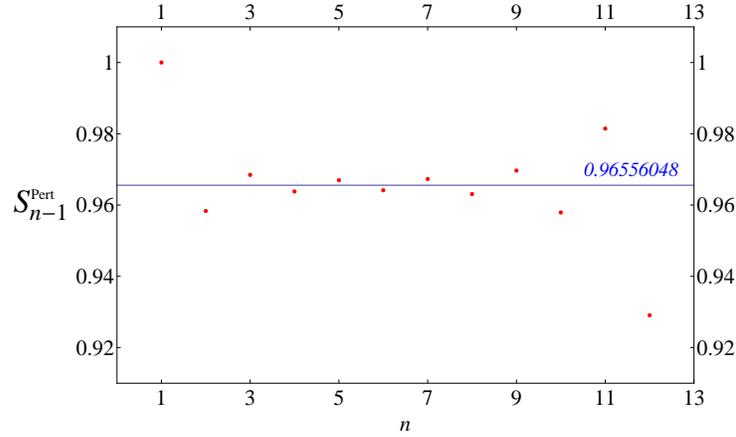}
\caption{$S^{\textrm{Pert}}_{n-1}$ ($n\in\{1,\dots,12\}$), for $\lambda=\frac{1}{3}$, compared to the exact result (\ref{numeriq}) (horizontal line).\label{fig2}}
\end{figure}

It is readily seen that the partial sums rapidly converge to a value around $0.965$ from which they f\/inally begin to diverge (see also Fig.~\ref{fig2}).

In fact, since (\ref{asymp}) is a divergent alternating asymptotic series, simple general interpretative considerations for its sum lead to the fact that, if we def\/ine the remainder $R_n$ so that
\[
S^{\textrm{Pert}}_\infty=S^{\textrm{Pert}}_{n-1}+R_n,
\]
one is led to (see Appendix A for details)
\[
|R_{n}|<|u_n|
\]
and
\[
|R_{n}|<|u_{n-1}|.
\]
We may conclude from these inequalities that the best truncation of the series (obtained by minimizing the remainder $R_n$) is theoretically obtained by truncating before or after the smallest term in magnitude.

Therefore, there are two ``best predictions'' from the perturbative expansion (\ref{asymp}) with $\lambda=\frac{1}{3}$, which only dif\/fer by their central value (see Table~\ref{table1}):
\begin{gather}
\label{pertpred}
Z(0)\big\vert^{\textrm{Pert-1}}_{\lambda=\frac{1}{3}}= 0.96696177\pm 0.00281979
\end{gather}
and
\[
Z(0)\big\vert^{\textrm{Pert-2}}_{\lambda=\frac{1}{3}}= 0.96414197\pm 0.00281979.
\]
It is clear that since (\ref{asymp}) is an alternating series, the exact value has to be between them, from which we conclude that
\begin{gather}
\label{pertpred3}
Z(0)\big\vert^{\textrm{Pert}}_{\lambda=\frac{1}{3}}= 0.96555187\pm 0.00140990
\end{gather}
so that the perturbative expansion (\ref{asymp}) leads to an already very good prediction (from the precision level viewpoint), of the order of 0.15$\%$.

In fact, from (\ref{vacuum}), one gets, with an 8 decimal places precision,
\begin{gather}
\label{numeriq}
Z(0)\big\vert_{\lambda=\frac{1}{3}}= 0.96556048.
\end{gather}
The central value in (\ref{pertpred3}) is therefore very close to the exact value, and corresponds actually to the standard Stieltjes approximative resummation formula for alternating divergent series, which reads  \cite[p.~402]{Dingle}
\begin{gather}
\label{Stieltjes}
Z(0)\simeq\sum_{k=0}^{\eta-1}u_k+\frac{1}{2}u_\eta,
\end{gather}
where $u_\eta$ is the term of least magnitude.

The point is that one can approach the ``right'' value (\ref{numeriq}) much closer than what perturbation theory does, by a ref\/ined asymptotic analysis that we present in the next sections where we obtain non-perturbative asymptotic improvements of the perturbative expansions~(\ref{asymp}) and~(\ref{intG2p}). At the f\/irst stage of this analysis, the improvement takes the form of so-called \textit{exponentially improved} asymptotic expansions \cite{Olver} but, the process being iterative, it will be possible to get more and more non-perturbative ref\/inements, in terms of \textit{hyperasymptotic} expansions of higher level.

\subsection{Exponential asymptotics: non-perturbative asymptotic improvement}\label{NP}

To go beyond perturbation theory, one could perform a Borel resummation of (\ref{asymp}) and, in this simple example, it works: one may reconstruct $Z(0)$ from its perturbative expansion by a Borel resummation, for ${\rm Re}\, \lambda >0$.

On the one hand, as we will see, the method we present here also allows to reconstruct $Z(0)$ from its perturbative expansion but, on the other hand, it gives another look at~$Z(0)$: as a~non-perturbative asymptotic method that reveals interesting ef\/fects that would be hidden in a~usual Borel resummation, like a~resurgence phenomenon.

\subsubsection{\label{NP1}Interpretation of the divergent perturbative expansion}

Numerically, we saw that the f\/irst few perturbative terms already do a very good work for the description of $Z(0)$, but that the divergent character of the perturbative expansion is unavoidable if one includes more and more terms. In order to solve this problem, we have to give a meaning to the tail of the perturbative expansion. With this in mind, we divide the perturbative expansion into two parts, following \cite[p.~406]{Dingle}:
\[
Z(0)\underset{\lambda\rightarrow 0}{\sim}\frac{1}{\sqrt{\pi}}\sum_{k=0}^{n-1}\frac{(-1)^k}{k!}
\frac{\Gamma\left(\frac{1}{2}+2k\right)}{(3!)^{k}}\lambda^k
+\frac{1}{\sqrt{\pi}}\sum_{k=n}^{+\infty}\frac{(-1)^k}{k!}\frac{\Gamma\left(\frac{1}{2}+2k\right)}{(3!)^{k}}\lambda^k.
\]
The f\/irst part is the perturbative contribution that one wants to keep and the second, the tail of the divergent series.

We are now going to perform formal manipulations that will be justif\/ied \textit{a posteriori}, in the next section.

First, it is convenient to use the duplication formula
\begin{gather}
\label{dupl}
\Gamma(2z)=\frac{1}{\sqrt{\pi}}2^{2z-1}\Gamma(z)\Gamma\left(z+\frac{1}{2}\right)
\end{gather}
to rewrite the tail as
\begin{gather}
\label{tail0}
\frac{1}{\sqrt{\pi}}\sum_{k=n}^{+\infty}\frac{(-1)^k}{k!}
\frac{\Gamma\left(\frac{1}{2}+2k\right)}{(3!)^{k}}\lambda^k=\frac{1}{\sqrt{2}\pi}
\sum_{k=n}^{+\infty}\frac{\Gamma\left(k+\frac{1}{4}\right)\Gamma\left(k+\frac{3}{4}\right)}{\Gamma(k+1)}
\left(\frac{-2\lambda}{3}\right)^k.
\end{gather}
Our main tool is the so-called inverse factorial expansion which may be obtained from Barnes' lemma~\cite{Olver2} (see also \cite[Chapter~2, Section~2.2]{Paris})
\begin{gather}
\label{inversefact}
\frac{\Gamma\left(k+\frac{1}{4}\right)\Gamma\left(k+\frac{3}{4}\right)}{\Gamma(k+1)}
=\sum_{j=0}^{m-1}(-1)^jA_j\Gamma(k-j)
+\frac{1}{2i\pi}\int_{c+m-i\infty}^{c+m+i\infty} ds \; f(s) \Gamma(k-s) ,
\end{gather}
where
\begin{gather}
f(s)=\frac{\Gamma\left(s+\frac{1}{4}\right)\Gamma\left(s+\frac{3}{4}\right)
\Gamma(-s)}{\Gamma\left(\frac{1}{4}\right)\Gamma\left(\frac{3}{4}\right)} ,
\qquad
\label{Aj}
A_j=\frac{\Gamma\left(j+\frac{1}{4}\right)\Gamma\left(j+\frac{3}{4}\right)}{j! \Gamma\left(\frac{1}{4}\right)\Gamma\left(\frac{3}{4}\right)},
\end{gather}
and the contour in the Mellin--Barnes integral on the right hand side is a straight line with $c\in\left]-1,0\right[$ (for $m=0$, $c\in\left]-\frac{1}{4},0\right[$ but we always take $m>0$). It is important to note that (\ref{inversefact}) is an exact equality\footnote{One may in fact also prove that it is an asymptotic equality \cite{Olver2}.} only for $c+m<k$ (since in our case $m$ and $k$ are integers, and since $\textrm{min}(k)=n$ and $c<0$, this is equivalent to $m\leq n$)\footnote{If one does not impose this constraint, i.e.\ if $m>n$, then one has the alternative expansion
\begin{gather*}
\frac{\Gamma\left(k+\frac{1}{4}\right)\Gamma\left(k+\frac{3}{4}\right)}{\Gamma(k+1)}
=\sum_{j=0}^{k-1}(-1)^jA_j\Gamma(k-j)  \\
{} +\sum_{j=k}^{m-1}\frac{(-1)^k}{(j-k)!}A_{j}\left[\psi(1+j-k) + \psi(j+1) - \psi\left(\frac{1}{4}+j\right)- \psi\left(\frac{3}{4}+j\right)\right]+\frac{1}{2i\pi}\int_{d+m-i\infty}^{d+m+i\infty} ds \; f(s) \Gamma(k-s) ,
\end{gather*}
which cannot be at the basis of the hyperasymptotic procedure that we show in this paper.}.

Inserting (\ref{inversefact}) in (\ref{tail0}), exchanging the two sums and the sum and integral signs, one f\/inds
\begin{gather*}
\frac{1}{\sqrt{\pi}}\sum_{k=n}^{+\infty}\frac{(-1)^k}{k!}\frac{\Gamma\left(\frac{1}{2}+2k\right)}{(3!)^{k}}\lambda^k  =\frac{1}{\sqrt{2}\pi}\sum_{j=0}^{m-1}(-1)^jA_j\sum_{k=n}^{+\infty}\Gamma(k-j)\left(\frac{-2\lambda}{3}\right)^k \\
\phantom{\frac{1}{\sqrt{\pi}}\sum_{k=n}^{+\infty}\frac{(-1)^k}{k!}\frac{\Gamma\left(\frac{1}{2}+2k\right)}{(3!)^{k}}\lambda^k =}{}
+\frac{1}{\sqrt{2}\pi}\frac{1}{2i\pi}\int_{c+m-i\infty}^{c+m+i\infty}ds\;f(s)\sum_{k=n}^{+\infty}\Gamma(k-s)\left(\frac{-2\lambda}{3}\right)^k.
\end{gather*}

Now we perform Borel resummations using the def\/inition~\cite[p.~406]{Dingle}
\[
\sum_{k=n}^{+\infty}\Gamma(k-j)\left(\frac{-2\lambda}{3}\right)^k
=\Gamma(n-j)\left(\frac{-2\lambda}{3}\right)^n\Lambda_{n-j-1}\left(\frac{3}{2\lambda}\right),
\]
where $\Lambda_\ell(x)$ is one of the so-called terminant functions \cite[p.~406]{Dingle}, def\/ined (when ${\rm Re}\, \ell >-1$ and $|\arg x|<\pi$)\footnote{\label{terminantfootnote}This terminant function can be extended to ${\rm Re}\, \ell \leq -1$ via, for instance, its absolute convergent expansion \cite[p.~407]{Dingle}.} as
\begin{gather}
\label{terminant}
\Lambda_\ell\left(x\right)=\frac{1}{\Gamma(\ell+1)}\int_0^\infty  dy\; \frac{y^{\ell}   e^{-y}}{1+\frac{y}{x}}  ,
\end{gather}
which can also be expressed as
\begin{gather}\label{renormalonUV}
\Lambda_\ell\left(x\right)=x^{\ell+1}  e^x  \Gamma\left(-\ell,x\right),
\end{gather}
where we recall that $\Gamma(a,x)$ is the incomplete gamma function def\/ined, for $|\arg x|<\pi$, as (see for instance \cite[p.~112]{Paris})
\begin{gather}\label{incompgammafunct}
\Gamma(a,x)=\int_x^\infty dy\ y^{a-1}e^{-y}.
\end{gather}
Notice that the expression (\ref{renormalonUV}) of the terminant coincides with the Borel integral of a general UV renormalon pole (see \cite[equation~(A.3), p.~35]{Beneke:2008ad}). It may also be seen as a Mellin transform.

At the end, one obtains (for $m\leq n$)
\begin{gather}
 \frac{1}{\sqrt{\pi}}\sum_{k=n}^{+\infty}\frac{(-1)^k}{k!}\frac{\Gamma\left(\frac{1}{2}+2k\right)}{(3!)^{k}}\lambda^k  =\frac{(-1)^n}{\sqrt{2}\pi}e^{\frac{3}{2\lambda}}\!\sum_{j=0}^{m-1}(-1)^{j}A_j\Gamma(n-j)
 \! \left(\frac{3}{2\lambda}\right)^{-j}\!\Gamma\!\left(-n+j+1,\frac{3}{2\lambda}\right) \!\nonumber\\
\label{MBB}
\qquad{}+\frac{(-1)^n}{\sqrt{2}\pi}e^{\frac{3}{2\lambda}}\frac{1}{2i\pi}
\int_{c+m-i\infty}^{c+m+i\infty}ds\left(\frac{3}{2\lambda}\right)^{-s}f(s)\Gamma(n-s)\Gamma
\left(-n+s+1,\frac{3}{2\lambda}\right).
\end{gather}
For the moment, this equation is valid in $|\arg \lambda|<\pi$ since the incomplete gamma functions which appear in the right hand side must be understood as derived from their integral representation~(\ref{incompgammafunct}).

We therefore conclude that, for $m\leq n$ and $|\arg \lambda|<\pi$,
\begin{gather}
Z(0)=\frac{1}{\sqrt{\pi}}\sum_{k=0}^{n-1}\frac{(-1)^k}{k!}
\frac{\Gamma\left(\frac{1}{2}+2k\right)}{(3!)^{k}}\lambda^k\nonumber\\
\phantom{Z(0)=}{} +\frac{(-1)^n}{\sqrt{2}\pi}e^{\frac{3}{2\lambda}}\sum_{j=0}^{m-1}(-1)^{j}A_j
\Gamma(n-j)\left(\frac{3}{2\lambda}\right)^{-j}\Gamma\left(-n+j+1,\frac{3}{2\lambda}\right)\nonumber\\
\phantom{Z(0)=}{}
 +\frac{(-1)^n}{\sqrt{2}\pi}\; e^{\frac{3}{2\lambda}} \frac{1}{2i\pi}\int_{c+m-i\infty}^{c+m+i\infty}ds\left(\frac{3}{2\lambda}\right)^{-s}   f(s)   \Gamma(n-s) \Gamma\left(-n+s+1,\frac{3}{2\lambda}\right) .\label{MBBZ0}
\end{gather}

The tail of the divergent perturbative series, (\ref{MBB}), has been rewritten as a partial sum, supplemented by a remainder integral written as a Mellin--Barnes representation. We have therefore converted an inf\/inite sum into a f\/inite sum plus a convergent integral and it will be proved in the Section~\ref{hyperMB} that the formal expression (\ref{MBBZ0}) is exact for any $n$ and $m$ as long as $n\geq m$ and constitutes what is called the f\/irst level of the hyperasymptotic expansion of $Z(0)$. Better than that, we will also prove in Section~\ref{NP-OTS} that the right hand side of (\ref{MBB}) is, at an optimal value of $n$ to be def\/ined later, exponentially suppressed with respect to $\lambda$, so that (\ref{MBB}) gives in fact, for this optimal value of $n$, the expression of a purely non-perturbative quantity.

Equation (\ref{MBBZ0}) is valid in a wider sector ($|\arg \lambda|<\pi$) than the usual Borel resummation which is valid for $|\arg \lambda|<\frac{\pi}{2}$. Moreover, we would like the reader to notice the similarity between the perturbative partial sum (f\/irst line of~(\ref{MBBZ0})) and the partial sum in the second line of~(\ref{MBBZ0}).
Indeed, choosing $n=5$ and $m=5$, we explicitly get from~(\ref{MBBZ0})
\begin{gather}
Z(0) =1-\frac{\lambda}{8}+\frac{35}{384}\lambda^2-\frac{385}{3072}\lambda^3+\frac{25025}{98304}\lambda^4 \nonumber\\
\phantom{Z(0) =}{} -\frac{1}{\sqrt{2}\pi}e^{\frac{3}{2\lambda}}\left\{\Gamma(5)\Gamma\left(-4,\frac{3}{2\lambda}\right)-\frac{1}{8}\lambda  \Gamma(4)\Gamma\left(-3,\frac{3}{2\lambda}\right) +\frac{35}{384}\lambda^2  \Gamma(3)\Gamma\left(-2,\frac{3}{2\lambda}\right) \right. \nonumber \\
 \left.
\phantom{Z(0) =}{}
 -\frac{385}{3072}\lambda^3  \Gamma(2)\Gamma\left(-1,\frac{3}{2\lambda}\right)+\frac{25025}{98304}
 \lambda^4\Gamma(1)\Gamma\left(0,\frac{3}{2\lambda}\right)\right\} \nonumber\\
\phantom{Z(0) =}{}
-\frac{1}{\sqrt{2}\pi}  e^{\frac{3}{2\lambda}}  \frac{1}{2i\pi}\int_{c+5-i\infty}^{c+5+i\infty}ds
\left(\frac{3}{2\lambda}\right)^{-s} f(s) \Gamma(5-s) \Gamma\left(-4+s,\frac{3}{2\lambda}\right)\label{MBBbis}.
\end{gather}
We see that the coef\/f\/icients of the second partial sum
are the same as the perturbative ones (up to Euler gamma functions that we wrote explicitly to emphasize the symmetry of the formula). We chose $n=5$ because, as we saw in (\ref{pertpred}), it is one of the two best orders for truncating the perturbative series for $\lambda=\frac{1}{3}$. Let us however underline that this interesting phenomenon is independent of the choice of $n$. Indeed, for example, an equivalent formula to (\ref{MBBbis}) is
\begin{gather}
Z(0)  = 1 -\frac{\lambda}{8}+\frac{35}{384}\lambda^2 \nonumber\\
\phantom{Z(0)  =}{}
-\frac{e^{\frac{3}{2\lambda}}}{\sqrt{2}\pi}\left\{\Gamma(3)\Gamma
\left(-2,\frac{3}{2\lambda}\right)-\frac{1}{8}\lambda \Gamma(2)\Gamma\left(-1,\frac{3}{2\lambda}\right)+\frac{35}{384}\lambda^2 \Gamma(1)\Gamma\left(0,\frac{3}{2\lambda}\right)\right\}\nonumber\\
\phantom{Z(0)  =}{}
-\frac{1}{\sqrt{2}\pi}e^{\frac{3}{2\lambda}}\frac{1}{2i\pi}\int_{c+3-i\infty}^{c+3+i\infty}ds
\left(\frac{3}{2\lambda}\right)^{-s}f(s)  \Gamma(3-s)  \Gamma\left(-2+s,\frac{3}{2\lambda}\right),\label{MBBbisbis}
\end{gather}
where one sees that although the terms in the second line of (\ref{MBBbisbis}) are not the same as those in the second line of (\ref{MBBbis}), their coef\/f\/icients are still equal to those of the perturbative contributions. This is a so-called resurgence phenomenon (see \cite[p.~271]{Paris}), which will also manifest itself at higher order in the non-perturbative asymptotic improvement process. It can be understood, at our present level, from the fact that the perturbative series (using the duplication formula and noting that $\sqrt{2}\pi=\Gamma(\frac{1}{4})\Gamma(\frac{3}{4})$) may be rewritten as{\samepage
\begin{gather}\label{asymp2}
Z(0)\underset{\lambda\rightarrow 0}{\sim} \frac{1}{\sqrt{\pi}}\sum_{k=0}^{+\infty}\frac{(-1)^k}{k!}
\frac{\Gamma\left(\frac{1}{2}+2k\right)}{(3!)^{k}}\lambda^k
=\sum_{k=0}^{+\infty}(-1)^kA_k\left(\frac{3}{2\lambda}\right)^{-k},
\end{gather}
which has the same coef\/f\/icients $A_k$ as those appearing in the second line of (\ref{MBBZ0}).}

An alternative expression for $Z(0)$ (equivalent to (\ref{MBBZ0}), except for its $\lambda$-domain of convergence, as we will see in the next section) is obtained by inserting in (\ref{terminant}) the Mellin--Barnes representation
\[
\frac{1}{1+\frac{y}{x}} = \frac{1}{2i\pi}\int_{d-i\infty}^{d+i\infty}   dt \; \left(\frac{y}{x}\right)^{-t}   \frac{\pi}{\sin(\pi t)}  ,
\]
where $d={\rm Re}\, t \in]0,1[$. Performing the $y$-integral one then f\/inds
\[
\Lambda_\ell\left(x\right)=\frac{1}{\Gamma(\ell+1)}\frac{1}{2i\pi}\int_{d-i\infty}^{d+i\infty}  dt \; \left(\frac{1}{x}\right)^{-t}   \frac{\pi}{\sin(\pi t)}\Gamma(l+1-t),
\]
with ${\rm Re}\,(l-t)>-1$, so that, at the end,
\begin{gather}
 Z(0)=\sum_{k=0}^{n-1}(-1)^kA_k\left(\frac{3}{2\lambda}\right)^{-k}  \nonumber \\
\phantom{Z(0)=}{}
 -\frac{1}{\sqrt{2}\pi}\sum_{j=0}^{m-1}(-1)^{j}A_j\frac{1}{2i\pi}\int_{d+n-i\infty}^{d+n+i\infty}dt  \left(\frac{3}{2\lambda}\right)^{-t} \frac{\pi}{\sin(\pi t)}\Gamma(t-j)\nonumber \\
\phantom{Z(0)=}{}
-\frac{1}{\sqrt{2}\pi}\frac{1}{2i\pi}\int_{c+m-i\infty}^{c+m+i\infty}ds  f(s)   \frac{1}{2i\pi}\int_{e+n-i\infty}^{e+n+i\infty}du  \left(\frac{3}{2\lambda}\right)^{-u}\frac{\pi}{\sin(\pi u)}\Gamma(u-s)\label{MBBinterpret}
\end{gather}
with $n\geq m$, $c\in]{-}1,0[$ and $d\in]{-}1,0[$. In the double Mellin--Barnes integral, $e\in]{-}\min(1,n-m-c),0[$.

For the $N$-point function case (with $N=2p\geq 2$), our formal strategy applied to (\ref{intG2p}) imposes an inverse factorial expansion with constraints on $n$ and $m$ dif\/ferent from those implied by (\ref{inversefact}), because of the $p$ dependence of the ratio of Gamma functions.

Indeed, the inverse factorial that we need is
\begin{gather}
\frac{\Gamma\left(k+\frac{p}{2}+\frac{1}{4}\right)\Gamma\left(k+\frac{p}{2}+\frac{3}{4}\right)}{\Gamma(k+1)}\nonumber\\
\qquad{}=\sum_{j=0}^{m-1}(-1)^jB_j^{(p)}\Gamma(k+p-j) +\frac{1}{2i\pi}\int_{c+m-i\infty}^{c+m+i\infty} ds \; g_p(s) \Gamma(k+p-s),\label{inversefact2}
\end{gather}
where the contour of the Mellin--Barnes integral in the right hand side is a straight line with $c$ a real number so that $c+m>-\frac{1}{4}+\frac{p}{2}$ and $c+m< n+p$, and
\[
g_p(s) = \frac{\Gamma\left(s - \frac{p}{2} +\frac{1}{4}\right)\Gamma\left(s - \frac{p}{2} +\frac{3}{4} \right)\Gamma(-s)}{\Gamma\left(\frac{1}{4} - \frac{p}{2}\right)\Gamma\left(\frac{3}{4} - \frac{p}{2}\right)}
\]
and
\[
B^{(p)}_j = \frac{\Gamma\left(j - \frac{p}{2}+\frac{1}{4}\right)\Gamma\left(j - \frac{p}{2} +\frac{3}{4} \right)}{j!   \Gamma\left(\frac{1}{4} - \frac{p}{2}\right)\Gamma\left(\frac{3}{4} - \frac{p}{2}\right)}.
\]
Contrary to (\ref{inversefact}), one cannot naively put $m=0$ in (\ref{inversefact2}). Indeed, Barnes' lemma has no fundamental strip in this case since $p>0$. To be valid here, Barnes' lemma needs a deformed contour of the Mellin--Barnes integral which separates the poles of $\Gamma\left(s - \frac{p}{2} +\frac{1}{4}\right)$ and $\Gamma\left(s - \frac{p}{2} +\frac{3}{4} \right)$ from those of $\Gamma(-s)\Gamma(k+p-s)$, while going from $-i\infty$  to $i\infty$.

Taking into account these facts one f\/inds, at the end,
\begin{gather}
\int_{-\infty}^{+\infty}d\phi\ \phi^{2p} e^{-\frac{1}{2}\phi^2-\frac{\lambda}{4!}\phi^4} =\frac{2^{2p}}{\sqrt{\pi}}  \Gamma\left(\frac{1}{4} + \frac{p}{2}\right)\Gamma\left(\frac{3}{4} + \frac{p}{2}\right)\sum_{k=0}^{n-1}(-1)^k B^{(-p)}_k   \left(\frac{3}{2\lambda}\right)^{-k}  \nonumber\\
\qquad{} + (-1)^n\frac{2^{2p}}{\sqrt{\pi}} \left(\frac{3}{2\lambda}\right)^pe^{\frac{3}{2\lambda}}  \sum_{j=0}^{m-1} (-1)^{j} B^{(p)}_j \Gamma(n+p-j) \left(\frac{3}{2\lambda}\right)^{-j}   \Gamma\left(-n-p+j+1,\frac{3}{2\lambda}\right)\nonumber\\
\qquad{}+(-1)^n\frac{2^{2p}}{\sqrt{\pi}} \left(\frac{3}{2\lambda}\right)^pe^{\frac{3}{2\lambda}}\int_{c+m-i\infty}^{c+m+i\infty} \frac{ds}{2i\pi} \left(\frac{3}{2\lambda}\right)^{-s} g_p(s) \Gamma(n+p-s)  \nonumber\\
\qquad{}\times   \Gamma\left(-n-p+s+1,\frac{3}{2\lambda}\right).
\label{MBBNpoint}
\end{gather}

\subsubsection[Mellin-Barnes hyperasymptotic theory]{Mellin--Barnes hyperasymptotic theory}\label{hyperMB}

As we shall see now, the results (\ref{MBBZ0}) and (\ref{MBBinterpret}) can be entirely justif\/ied from the modern point of view of Mellin--Barnes hyperasymptotic theory  \cite[Chapter~6]{Paris} if one does not anymore ignore, contrary to what we did since the beginning, the integral representation of~$Z(0)$. Indeed, after a computation of the asymptotic expansion of $Z(0)$ from its Mellin--Barnes representation it is possible to obtain, from an expansion of the Mellin--Barnes asymptotic remainder integral, the hyperasymptotic expansion of~$Z(0)$ at f\/irst hyperasymptotic level. As we will see the results of this approach will match those of Section~\ref{NP1} and therefore provide a proof that the formal strategy that we proposed is correct.

After this, we will show how an expansion of the f\/irst level hyperasymptotic remainder integral (third line of (\ref{MBBinterpret})) leads to the second hyperasymptotic level. The hyperasymptotic procedure may then be iterated and one can obtain an hyperasymptotic expansion at an arbitrary hyperasymptotic level.

From the study of the Mellin--Barnes integrals involved in this framework we will conclude that~(\ref{MBBinterpret}) is valid for a wider range of phase than the complex $\lambda$-plane with a cut on the negative real axis. In particular it will be proved to be valid on the Stokes lines def\/ined by $|\arg \lambda|=\pi$.

Let us begin the calculations.
Using the Mellin--Barnes representation
\[
e^{-\frac{\lambda}{4!}\phi^4}=\frac{1}{2i\pi}\int_{c-i\infty}^{c+i\infty}ds
\left(\frac{\lambda}{4!}\phi^4\right)^{-s}\Gamma(s),
\]
which is valid in the semi-inf\/inite fundamental strip def\/ined by $c={\rm Re}\, s\in]0,+\infty[$ (the right half-complex $s$-plane) and for $|\arg \lambda|<\frac{\pi}{2}$, one obtains, from (\ref{vacuum}),
\[
Z(0)=\frac{1}{\sqrt{2\pi}2i\pi}\int_{c-i\infty}^{c+i\infty}ds\left(\frac{\lambda}{4!}\right)^{-s}
\Gamma(s)\int_{-\infty}^{+\infty}d\phi\ \phi^{-4s}e^{-\frac{1}{2}\phi^2}.
\]
The $\phi$ integral can be computed and we f\/ind
\begin{gather}
\label{MBZ}
Z(0)=\frac{1}{2i\pi\sqrt{\pi}}\int_{c-i\infty}^{c+i\infty}ds\; \lambda^{-s}\frac{\Gamma(s)\Gamma\left(\frac{1}{2}-2s\right)}{(3!)^{-s}},
\end{gather}
where $c={\rm Re} \, s\in]0,\frac{1}{4}[$ so that the $\phi$ integral \textit{reduced} the semi-inf\/inite fundamental strip to a~f\/inite one. Moreover our Mellin--Barnes representation \textit{increases} the $\lambda$-domain of validity\footnote{\label{conv}The $\lambda$-domain of convergence of a Mellin--Barnes integral \textit{of the type} (\ref{MBZ}) is given by $|\arg \lambda|<\frac{\pi}{2}(N_n-N_d)$, where $N_n$ is the number of Gamma functions in the numerator of the integrand and $N_d$ the number of Gamma functions in the denominator, taking into account the multiplicity of each Gamma function, the multiplicity of a Gamma function being def\/ined as the absolute value of the number multiplying the variable in the argument of this function  \cite[Chapter~2, Section~4]{Paris}. For instance, in the case of (\ref{MBZ}), we have $N_n=1+|-2|=3$ and $N_d=0$.} of~$Z(0)$ to $|\arg \lambda|<\frac{3\pi}{2}$, therefore (\ref{MBZ}) provides an analytic continuation of (\ref{vacuum}) on the two sheets which are adjacent to the principal sheet of its Riemann surface (we recall that (\ref{vacuum}) is def\/ined in the right half complex plane only, i.e.\ for $|\arg \lambda|<\frac{\pi}{2}$).

It is of course possible to apply the whole hyperasymptotic machinery directly on (\ref{MBZ}), in the $|\lambda|\rightarrow 0$ limit\footnote{This is what we described in detail in the initial version of this paper. We thank one of the referees for his suggestion to use equation~(\ref{confluentZ0}) in order to make the presentation more concise.}. However, it is more convenient to notice that, from the def\/inition of the conf\/luent hypergeometric function $U(a,a-b+1,z)$ \cite[p.~506]{AS} and with the help of the duplication formula (\ref{dupl}), one has
\begin{gather}
\label{confluentZ0}
Z(0)=\left(\frac{3}{2\lambda}\right)^{\frac{1}{4}}U\left(\frac{1}{4},\frac{1}{2},\frac{3}{2\lambda}\right).
\end{gather}
Indeed, with this representation, it is now possible to apply the following hyperasymptotic theorem \cite[p.~270]{Paris}:

\begin{theorem}\label{hypertheorem}
The expansion of the confluent hypergeometric function $U(a, a-b+1,z)$ for $|z|\rightarrow\infty$  is given by
\begin{gather}\label{UasympRzero}
z^aU(a,a-b+1,z)= \sum_{k=0}^{n_0-1}(-1)^k\frac{\Gamma(k+a)\Gamma(k+b)}{k!\ \Gamma(a)\Gamma(b)}z^{-k}+R_0(z),
\end{gather}
where the remainder $R_0(z)$ in the truncation of the Poincar\'e asymptotic series after $n_0$ terms possesses the hyperasymptotic expansion
\begin{gather}\label{RhyperU}
R_0(z)=\frac{2\pi(-1)^{n_0}}{\Gamma(a)\Gamma(b)}z^\theta e^z\sum_{j=0}^{m-1}\Delta_j(a,b)\sum_{k=0}^{n_{j+1}-1}(-1)^kA_k^{(j)}{\bf T}_j(z,k)+R_m(z),
\end{gather}
for $m=1,2,\dots$, with $\theta=a+b-1$,
\begin{gather}
 \Delta_j(a,b)= \left\{
    \begin{array}{ll}
      (\sin(\pi a)\sin(\pi b))^{\frac{j}{2}}\quad & (j \ \mbox{even}),\\
     -\dfrac{\pi(\sin(\pi a)\sin(\pi b))^{\frac{(j-1)}{2}}}{\Gamma(1-a)\Gamma(1-b)}\quad & (j \ \mbox{odd}),
    \end{array}
  \right.
\nonumber\\
\label{ak}
 A_k^{(j)}= \left\{
    \begin{array}{ll}
      \dfrac{\Gamma(k+1-a)\Gamma(k+1-b)}{k!  \Gamma(1-a)\Gamma(1-b)}\quad &  (j \ \mbox{even}),\\
     \dfrac{\Gamma(k+a)\Gamma(k+b)}{k!  \Gamma(a)\Gamma(b)}\quad & (j \ \mbox{odd})
    \end{array}
  \right.
\end{gather}
and $($for $j\geq1)$
\[
{\bf T}_j(z,k)=\frac{1}{2i\pi}\int_{L(n_j)} ds_j\frac{\Gamma(s_j+(-1)^j\theta-k)}{\sin(\pi s_j)}{\bf T}_{j-1}(z,s_j),
\]
where $L(n_j)$ is the path of integration parallel to the imaginary axis of the integration variable~$s_j$ $({\rm Re}\,(s_j)=c_j+n_j$, with $-1<c_j<0)$ and
\[
{\bf T}_0(z,k)=z^{-k}T_{n_0+\theta-k}(z),
\]
with{\samepage
\[
T_{n_0+\theta-k}(z)=(-1)^{n_0+1}\frac{z^{-\theta+k}}{4i\pi}e^{-z}\int_{c+n_0-i\infty}^{c+n_0+i\infty}ds_0\ z^{-s_0}\frac{\Gamma(s_0+\theta-k)}{\sin(\pi s_0)},
\]
where $-1<c<0$.}

Moreover, the remainder $R_m(z)$ in \eqref{RhyperU} is defined by
\[
R_m(z)=\frac{2\pi(-1)^{n_0}}{\Gamma(a)\Gamma(b)}z^\theta e^z\hat R_m(z),
\]
where
\begin{gather}\label{***}
\hat R_m(z)=\frac{\Delta_m(a,b)}{2i\pi}\int_{L(n_m)} ds_m\frac{\Gamma(s_m+a_m)\Gamma(s_m+b_m)}{\Gamma(s_m+1)\sin(\pi s_m)}{\bf T}_{m-1}(z,s_m),
\\
%\label{am}
\nonumber
a_m= \left\{
    \begin{array}{ll}
      a\quad & (m \ \mbox{even}),\\
     1-a\quad & (m \ \mbox{odd})
    \end{array}
  \right.
\end{gather}
and
\begin{gather*}%\label{bm}
b_m= \left\{
    \begin{array}{ll}
     b\quad & (m \ \mbox{even}), \\
     1-b\quad & (m \ \mbox{odd}).
    \end{array}
  \right.
\end{gather*}
For $j\geq1$ the $n_j$ are integers such that
\[
\left\{
    \begin{array}{lll}
      n_j+c_j>\max({\rm Re}\, (a-1), {\rm Re}\, (b-1)), \qquad &  n_j+c_j<{\rm Re}\, (s_{j-1}+\theta)\quad &  (j \ \mbox{odd}),\\
     n_j+c_j>\max({\rm Re}\, (-a),  {\rm Re}\, (-b)),\qquad  & n_j+c_j<{\rm Re}\, (s_{j-1}-\theta)\quad & (j \ \mbox{even}).
    \end{array}
  \right.
\]
\end{theorem}

Let us apply this theorem to $Z(0)$, i.e.\ for $a=\frac{1}{4}$, $b=\frac{3}{4}$ and $z=\frac{3}{2\lambda}$.
For this special case $\theta=0$ and, therefore, there is no dif\/ference
between even and odd cases in (\ref{ak}) and in the ratio of gamma functions
in the integrand of \eqref{***} so that
% For this special case $\theta=0$ and, more interestingly, there is no dif\/ference between even and odd cases in equations~(\ref{ak}), (\ref{am}) and (\ref{bm}) so that
 one obtains, for $m=1$ (the f\/irst hyperasymptotic level) and after simplif\/ication,
\begin{gather}
\nonumber
Z(0)=\frac{1}{\sqrt{\pi}}\sum_{k=0}^{n_0-1}\frac{(-1)^k}{k!}
\frac{\Gamma\left(\frac{1}{2}+2k\right)}{(3!)^{k}}\lambda^k\\
\phantom{Z(0)=}{}
-\frac{1}{\sqrt{2}\pi}\sum_{k=0}^{n_1-1}(-1)^{k}A_k\frac{1}{2i\pi}
\int_{c+n_0-i\infty}^{c+n_0+i\infty}ds\left(\frac{3}{2\lambda}\right)^{-s} \frac{\pi}{\sin(\pi s)}\Gamma(s-k) \nonumber\\
\phantom{Z(0)=}{}-\frac{1}{\sqrt{2}\pi}  \frac{1}{2i\pi}\!\int_{c_1+n_1-i\infty}^{c_1+n_1+i\infty}\!ds_1 \; f(s_1)   \frac{1}{2i\pi}\!\int_{c+n_0-i\infty}^{c+n_0+i\infty}\!ds \left(\frac{3}{2\lambda}\right)^{-s} \frac{\pi}{\sin(\pi s)}\Gamma(s-s_1)\label{hypertheorem1}
\end{gather}
with $n_1+c_1<n_0+c$. The $A_k$ are those of equation~(\ref{Aj}).

Now, this last formula is exactly (\ref{MBBinterpret}). Of course, one may also show that (\ref{hypertheorem1}) is equivalent to (\ref{MBBZ0}) if one replaces the incomplete gamma function in the second line of (\ref{MBBZ0}) by its Mellin--Barnes representation
\begin{gather}
\Gamma\left(-n+x+1,\frac{3}{2\lambda}\right) \nonumber \\
\qquad{}=\frac{-1}{\Gamma(n-x)}\left(\frac{3}{2\lambda}\right)^{-n+x}e^{-\frac{3}{2\lambda}}
\frac{1}{2i\pi}\int_{d-i\infty}^{d+i\infty}dt\ \left(\frac{3}{2\lambda}\right)^{-t}\frac{\pi}{\sin(\pi t)}
\Gamma(t+n-x) ,\label{incompleteMB}
\end{gather}
with $d={\rm Re}\, t\in]{-}1,0[$, in the third line of (\ref{MBBZ0}) by the same formula but with $d={\rm Re}\, t\in]{-}\min(1,n-m-c),0[$ and if one performs the change of variable $t'=t+n$.

We have therefore proved that the interpretation of the tail of the divergent perturbative expansion given by (\ref{MBB}) is correct and that (\ref{MBBZ0}), or (\ref{MBBinterpret}), constitute the f\/irst level of the hyperasymptotic expansion of $Z(0)$.

An important remark is that, written in terms of Mellin--Barnes representations, our expansion (\ref{MBBZ0}) obtained initially for values of $\lambda$ in the complex plane with a cut on the negative real axis, is now automatically valid in the wider sector $|\arg \lambda |<\frac{3\pi}{2}$ (see footnote \ref{conv}) where it also gives much more precise results than perturbation theory.%(\ref{asympMB}).

Let us illustrate this fact by computing $Z(0)\vert_{\lambda=\frac{1}{3}e^{i\pi}}$.
From (\ref{MBZ}), after performing the change of variable $t=-i(s-c)$, one gets numerically, with an 8 decimal places precision,
\begin{gather}
\label{precisionIm}
Z(0)\big\vert_{\lambda=\frac{1}{3}e^{i\pi}}=1.05995021-0.00758472i.
\end{gather}
But from (\ref{asymp2}) \textit{which is valid}\footnote{In deed, the asymptotic expansion of (\ref{MBZ}) when $|\lambda|\rightarrow 0$ leads to (\ref{asymp2}) for $|\arg \lambda|<\frac{3\pi}{2}$.} \textit{for} $\lambda=\frac{1}{3}e^{i\pi}$, the best prediction is\footnote{Equation (\ref{ImNum}) is obtained in the same way as (\ref{pertpred3}).}
\begin{gather}
\label{ImNum}
Z(0)\big\vert_{\lambda=\frac{1}{3}e^{i\pi}}=1.06098837\pm 0.00140990.
\end{gather}
so that the purely perturbative approach of course completely misses the imaginary part.

In fact, imaginary contributions appear from the second sum of (\ref{MBBinterpret}) (or of (\ref{MBBZ0}) with the insertion of (\ref{incompleteMB})). Indeed,
truncating the perturbative series after the f\/ifth term and including contributions of the f\/irst two terms of the second line of (\ref{MBBinterpret}) we get%(in OTS2,which is the best optimal truncation scheme at this level) we get
\begin{gather}
\label{ImNumExp}
Z(0)\big\vert_{\lambda=\frac{1}{3}e^{i\pi}}=1.05990083- 0.00752794i.
\end{gather}
We see that even without taking into account the remainder integral (third line of (\ref{MBBinterpret})), whose contribution would make us fall on the exact result (\ref{precisionIm}), we obtain a very good description of~$Z(0)$ for values of~$\lambda$ for which (\ref{asymp2}) gives a poor approximation.

Now it is straightforward to get the hyperasymptotic expansion of $Z(0)$ at the second level from the battle-horse formula (\ref{MBBinterpret}).
It will appear from an expansion of the double Mellin--Barnes remainder integral (third line of (\ref{MBBinterpret})) based on the inverse factorial expansion.

In deed, noticing that
\begin{gather}
\nonumber
f(s)  = -\frac{\Gamma\left(s+\frac{1}{4}\right)\Gamma\left(s+\frac{3}{4} \right)}{\Gamma\left(\frac{1}{4}\right)\Gamma\left(\frac{3}{4} \right)\Gamma(s+1)}   \frac{\pi}{\sin (\pi s)}
\\
\phantom{f(s)}{}
=-\frac{\pi}{\sin (\pi s)} \frac{1}{\Gamma\left(\frac{1}{4}\right)\Gamma\left(\frac{3}{4}\right)}\left[ \sum_{l=0}^{m'-1}(-1)^l A_l\Gamma(s-l) + \int _{h+m'-i\infty}^{h+m'+i\infty} \frac{dt}{2i\pi}   f(t)   \Gamma(s-t)\right],
\label{resurgInvFact1}
\end{gather}
where $h+m'<{\rm Re}\, s$ and $h\in]0,1[$, the f\/inal result reads
\begin{gather}
Z(0)= \sum_{k=0}^{n-1}(-1)^kA_k\left(\frac{3}{2\lambda}\right)^{-k} \!
 -\frac{1}{\sqrt{2}\pi}\sum_{j=0}^{m-1}(-1)^{j}A_j\frac{1}{2i\pi}\int_{d+n-i\infty}^{d+n+i\infty}dt  \left(\frac{3}{2\lambda}\right)^{-t}\frac{\pi}{\sin(\pi t)}\Gamma(t-j) \nonumber \\
\phantom{Z(0)=}{} +\frac{1}{\left(\sqrt{2}\pi\right)^2}\sum_{l=0}^{m'-1}(-1)^{l}A_l\left(\frac{1}{2i\pi}\right)^2\int_{e+n-i\infty}^{e+n+i\infty}du\ \left(\frac{3}{2\lambda}\right)^{-u}\frac{\pi}{\sin(\pi u)} \nonumber \\
\phantom{Z(0)=}{} \quad\times\int_{c+m-i\infty}^{c+m+i\infty}ds\ \Gamma(s-l)\frac{\pi}{\sin(\pi s)}\Gamma(u-s)
\nonumber \\
\phantom{Z(0)=}{} +\frac{1}{\left(\sqrt{2}\pi\right)^2}\left(\frac{1}{2i\pi}\right)^3\int_{e+n-i\infty}^{e+n+i\infty}du\left(\frac{3}{2\lambda}\right)^{-u}\frac{\pi}{\sin(\pi u)}\int_{c+m-i\infty}^{c+m+i\infty}ds\;\frac{\pi}{\sin(\pi s)} \Gamma(u-s)   \nonumber \\
\phantom{Z(0)=}{} \quad\times\int_{h+m'-i\infty}^{h+m'+i\infty}dt\ f(t)\Gamma(s-t) \label{MBB4}.
\end{gather}
The constraint $h+m'<{\rm Re}\, s $ implies $h+m'<c+m$ in (\ref{MBB4}).

We have now three interwoven partial sums where the third one is expressed in terms of the so-called hyperterminants, here def\/ined as double Mellin--Barnes integrals. Equation (\ref{MBB4}) of course matches the expression obtained from the hyperasymptotic theorem for $m=2$.

For the $N$-point function case, we have, at second hyperasymptotic level\footnote{We give here the hyperasymptotic expansion in an alternative form where (hyper)terminants are written in terms of (integrals of) incomplete gamma functions.},
\begin{gather*}
 \int_{-\infty}^{+\infty}d\phi\ \phi^{2p} e^{-\frac{1}{2}\phi^2-\frac{\lambda}{4!}\phi^4} =\frac{2^{2p}}{\sqrt{\pi}}  \Gamma\left(\frac{1}{4} + \frac{p}{2}\right)\Gamma\left(\frac{3}{4} + \frac{p}{2}\right)\sum_{k=0}^{n-1}(-1)^k B^{(-p)}_k   \left(\frac{3}{2\lambda}\right)^{-k}  \nonumber\\
\qquad{} + (-1)^n\frac{2^{2p}}{\sqrt{\pi}} \left(\frac{3}{2\lambda}\right)^pe^{\frac{3}{2\lambda}}  \sum_{j=0}^{m-1} (-1)^{j} B^{(p)}_j \Gamma(n+p-j) \left(\frac{3}{2\lambda}\right)^{-j}   \Gamma\left(-n-p+j+1,\frac{3}{2\lambda}\right)\nonumber\\
\qquad{} -(-1)^n\frac{2^{2p}}{\sqrt{\pi}} \left(\frac{3}{2\lambda}\right)^pe^{\frac{3}{2\lambda}}
\frac{1}{\Gamma\left(\frac{1}{4}-\frac{p}{2}\right)\Gamma\left(\frac{3}{4}-\frac{p}{2}\right)}\sum_{l=0}^{m'-1} (-1)^{l} B^{(-p)}_l  \nonumber\\
\qquad\quad{} \times \int_{c+m-i\infty}^{c+m+i\infty} \frac{ds}{2i\pi}  \left(\frac{3}{2\lambda}\right)^{-s} \frac{\pi}{\sin(\pi s)}  \Gamma(s-p-l) \Gamma(n+p-s)  \Gamma\left(-n-p+s+1,\frac{3}{2\lambda}\right) \nonumber\\
\qquad{} -(-1)^n\frac{2^{2p}}{\sqrt{\pi}} \left(\frac{3}{2\lambda}\right)^pe^{\frac{3}{2\lambda}} \frac{1}{\Gamma\left(\frac{1}{4}-\frac{p}{2}\right)\Gamma\left(\frac{3}{4}-\frac{p}{2}\right)}\nonumber\\
\qquad\quad{} \times \int_{c+m-i\infty}^{c+m+i\infty} \frac{ds}{2i\pi}  \left(\frac{3}{2\lambda}\right)^{-s}\frac{\pi}{\sin(\pi s)} \Gamma(n+p-s)  \Gamma\left(-n-p+s+1,\frac{3}{2\lambda}\right) \nonumber\\
\qquad\quad{} \times\int_{c+m'-i\infty}^{c+m'+i\infty} \frac{dt}{2i\pi}   g_{-p}(t)   \Gamma(s-p-t)\nonumber
\end{gather*}
with
\begin{gather}
g_p(s)  = -\frac{\Gamma\left(s+\frac{1}{4} - \frac{p}{2} \right)\Gamma\left(s+\frac{3}{4} - \frac{p}{2}  \right)}{\Gamma\left(\frac{1}{4} - \frac{p}{2}\right)\Gamma\left(\frac{3}{4} - \frac{p}{2}\right)\Gamma(s+1)}   \frac{\pi}{\sin (\pi s)}   =-\frac{\pi}{\sin (\pi s)} \frac{1}{\Gamma\left(\frac{1}{4} - \frac{p}{2}\right)\Gamma\left(\frac{3}{4} - \frac{p}{2}\right)}\nonumber \\
\phantom{g_p(s)=}{} \times\left[ \sum_{l=0}^{m'-1}(-1)^l B^{(-p)}_l\Gamma(s-p-l) + \int_{c+m'-i\infty}^{c+m'+i\infty} \frac{dt}{2i\pi}  g_{-p}(t)  \Gamma(s-p-t)\right].
\label{resurgInvFact2}
\end{gather}

The resurgence phenomenon is clearly apparent in (\ref{MBB4}) since the coef\/f\/icient $A_k$ appears at each hyperasymptotic order. One may however notice that for the $N$-point functions case, resurgence manifests itself in a dif\/ferent way: the coef\/f\/icients $B_j^{(p)}$ and $B_j^{(-p)}$ appear alternately at successive levels in the expansion.

This dif\/ference comes from the fact that for $Z(0)$ the function $f(s)$ in the left hand side of (\ref{resurgInvFact1}) appears also under the integral sign in the right hand side, whereas for the $N$-point functions the function $g_p(s)$ in the left hand side of (\ref{resurgInvFact2}) appears as $g_{-p}(s)$ under the integral sign in the right hand side (we also saw that for $Z(0)$ there is no dif\/ference between even and odd cases in equations~(\ref{ak})).

Thanks to these interesting symmetries that are ref\/lected by the resurgence phenomenon, one may deduce that in our cases of study, hyperasymptotic expansions at arbitrary hyperasymptotic levels may be obtained in a straightforward way and it is therefore possible to rederive the results of Theorem~\ref{hypertheorem} from the formal strategy that we proposed in Section~\ref{NP1}.

\subsubsection{Non-perturbative asymptotic improvement of perturbation theory:\\ optimal truncation schemes}\label{NP-OTS}

In Section~\ref{NP1}, we formally gave a meaning to the tail of the divergent perturbative expansion of $Z(0)$ and of an arbitrary $N$-point function, in terms of a partial sum added to a Mellin--Barnes remainder integral. This allowed us to obtain equations~(\ref{MBBZ0}), (\ref{MBBinterpret}) and (\ref{MBBNpoint}) which are in fact hyperasymptotic expansions at f\/irst hyperasymptotic level. Our interpretation of the tail has been rigorously justif\/ied in Section~\ref{hyperMB} and we saw in (\ref{ImNumExp}) that numerically the partial sum in the second line of (\ref{MBBinterpret}) gives contributions that perturbation theory (\ref{asymp2}) does not see (in particular an imaginary part, see (\ref{ImNum})), so that of course there is something beyond perturbation theory in the second line of~(\ref{MBBinterpret}). In this section we explain the non-perturbative nature of these results. The crucial point is that ``\textit{by allowing the number of terms in an asymptotic expansion to depend on the asymptotic variable, it is possible to obtain an error term that is exponentially small as the asymptotic variable tends to its limit}''~\cite{Olver}. In other words, there is an optimal value of the truncation index of the perturbative series for which the remainder of the expansion is a~non-perturbative quantity and in our example this optimal value is obtained by minimizing the remainder with $n$ scaling like $\frac{1}{|\lambda|}$. This will def\/ine optimal truncation schemes. Let us see this in more detail.

From the integral representation (\ref{MBZ}) it is straightforward to prove that the remainder of~(\ref{UasympRzero}) in the case of~$Z(0)$ may be written as\footnote{One may also obtain (\ref{reste2}) from the formal strategy of Section~\ref{NP1} by putting $m=0$ in (\ref{inversefact}), i.e.\ by using Barnes' lemma instead of the inverse factorial expansion.}:
\begin{gather}
\label{reste2}
\mathsf{R}_0\left(\frac{3}{2\lambda}\right)
=-\frac{1}{2i\pi\sqrt{2}\pi}\int_{d+n-i\infty}^{d+n+i\infty}dt\left(\frac{3}{2\lambda}\right)^{-t}\frac{\pi}{\sin(\pi t)}\frac{\Gamma\left(t+\frac{1}{4}\right)\Gamma\left(t+\frac{3}{4}\right)}{\Gamma(t+1)},
\end{gather}
where $d={\rm Re}\, t \in]{-}1,0[$.

Now, let us consider the integral
\begin{gather}
\label{reste3}
\mathsf{I}(z)=\frac{1}{2i\pi}\int_{c+n-i\infty}^{c+n+i\infty}dt\ z^{-t}\frac{\pi}{\sin(\pi t)}\frac{\Gamma\left(t+\alpha\right)\Gamma\left(t+\beta\right)}{\Gamma(t+1)},
\end{gather}
where $c={\rm Re}\, t\in]{-}1,0[$ and $n$ is a positive integer so that the path of integration in (\ref{reste3}) lies to the right of the poles of $\Gamma\left(t+\alpha\right)\Gamma\left(t+\beta\right)$. Then, as $n\rightarrow\infty$, we have, for $|\arg(z)|\leq\pi$ and $\omega=\alpha+\beta-1$,
\[
|\mathsf{I}(z)|=|z|^{-n-c}\mathcal{O}(e^{-n}n^{n+\omega+c}).
\]
This is the result of a lemma taken from \cite[Chapter~2, Section~5]{Paris}, where cases dealing with more general integrals than (\ref{reste3}) are also considered. One deduces from it that, since (\ref{reste2}) fulf\/ills the required condition,
\[
\left|\mathsf{R}_0\left(\frac{3}{2\lambda}\right)\right|
=\left|\frac{3}{2\lambda}\right|^{-n-d}\mathcal{O}\big(e^{-n}n^{n+d}\big).
\]
Now if, when $|\lambda|$ is small, we choose $n=\frac{a_0}{|\lambda|}+b_0$ (where $a_0>0$ and $b_0$ is bounded), then
\[
\left|\left.\mathsf{R}_0\left(\frac{3}{2\lambda}\right)\right\vert_{n=\frac{a_0}{|\lambda|}
+b_0}\right|=\mathcal{O}\left(\left[e^{-a_0}  \left(\frac{2a_0}{3}\right)^{a_0}\right]^{\frac{1}{|\lambda|}}\right).
\]
Minimizing the remainder then implies that $a_0=\frac{3}{2}$ and one f\/inally has
\[
\left|\left.\mathsf{R}_0\left(\frac{3}{2\lambda}\right)\right\vert_{n=\frac{3}{|2\lambda|}+b_0}\right|
=\mathcal{O}\big(e^{-\frac{3}{2|\lambda|}}\big).
%\label{Rsuper}
\]
If one chooses for instance $b_0$ such that $\frac{3}{2|\lambda|}+b_0=[\frac{3}{2|\lambda|}]$ (the integer part of $\frac{3}{2|\lambda|}$), then one obtains
\begin{gather}
\label{superasymp}
Z(0)=\frac{1}{\sqrt{\pi}}\sum_{k=0}^{[\frac{3}{2|\lambda|}]-1}\frac{(-1)^k}{k!}
\frac{\Gamma\left(\frac{1}{2}+2k\right)}{(3!)^{k}}\lambda^k+\mathcal{O}\big(e^{-\frac{3}{2|\lambda|}}\big),
\end{gather}
as $|\lambda|\rightarrow 0$ in $\left |\arg\left(\frac{3}{2\lambda}\right)\right |\leq\pi$.

For this optimal value of the perturbative truncation index the remainder is therefore non-per\-turbative (exponentially suppressed with respect to the coupling constant $|\lambda|$). Equa\-tion~(\ref{superasymp}) is called the superasymptotic expansion of~$Z(0)$, or optimally truncated hyperasymptotic expansion at zeroth level.

Now, at f\/irst hyperasymptotic level, we saw that the remainder (\ref{reste2}) has been expanded and a similar analysis may be carried out to deal with the remainder $\mathsf{R}_1$ of this expansion. Indeed, from (\ref{MBBinterpret}) one has
\begin{gather}
\mathsf{R}_0\left(\frac{3}{2\lambda}\right) =-\frac{1}{\sqrt{2}\pi}\!\sum_{j=0}^{m-1}(-1)^jA_j\frac{1}{2i\pi}\!\int_{d+n-i\infty}^{d+n+i\infty}\!\!
dt\left(\frac{3}{2\lambda}\right)^{-t}\!\!\frac{\pi}{\sin(\pi t)}\Gamma(t-j)
%\nonumber\\ \phantom{\mathsf{R}_0\left(\frac{3}{2\lambda}\right)=}{}
+\mathsf{R}_{1}\left(\frac{3}{2\lambda}\right)\! ,\!
\label{hyperRnm}
\end{gather}
with
\[
\mathsf{R}_1\left(\frac{3}{2\lambda}\right)  \doteq-\frac{1}{\sqrt{2}\pi}\frac{1}{2i\pi}\int_{e+n-i\infty}^{e+n+i\infty}du
\left(\frac{3}{2\lambda}\right)^{-u}\frac{\pi}{\sin(\pi u)}
\frac{1}{2i\pi}\!\!\int_{c+m-i\infty}^{c+m+i\infty}ds f(s) \Gamma(u-s),
\]
where we recall that $f(s)=\frac{\Gamma\left(s+\frac{1}{4}\right)
\Gamma\left(s+\frac{3}{4}\right)\Gamma(-s)}{\Gamma\left(\frac{1}{4}\right)\Gamma\left(\frac{3}{4}\right)}$.

From the same lemma as before and another one that may be found in~\cite[Chapter 2, Section~5]{Paris} one f\/inds that when $n$ and $m$ become inf\/inite (with $c=d$ and $n\geq m$)
\[
\left|\mathsf{R}_1\left(\frac{3}{2\lambda}\right)\right| =\mathcal{O}\left(\left|\frac{3}{2\lambda}\right|^{-n-c} e^{-n}(n-m)^{n-m}m^{m+c-\frac{1}{2}}\right).
\]
Since we found that $n=\frac{3}{2|\lambda|}+b_0$ was the optimal value for the truncation index of the perturbative series, one may keep this value for $n$ and choose $m=\frac{a_1}{|\lambda|}+b_1$ (where $a_1>0$ and $b_1$ are bounded), so that
\[
\left|\left.\mathsf{R}_1\left(\frac{3}{2\lambda}\right)\right\vert_{n=\frac{3}{2|\lambda|}+b_0,\ m=\frac{a_1}{|\lambda|}+b_1}\right|=\mathcal{O}\left(\frac{1}{|\lambda|^{-\frac{1}{2}}}
\left(e^{-\frac{3}{2}}a_1^{a_1}\left(\frac{3}{2}\right)^{-\frac{3}{2}}
\left(\frac{3}{2}-a_1\right)^{\frac{3}{2}-a_1}\right)^{\frac{1}{|\lambda|}}\right).
\]
as $|\lambda|\rightarrow 0$ in $\left |\arg\left(\frac{3}{2\lambda}\right)\right |\leq\pi$.

Minimizing the remainder implies $a_1=\frac{3}{4}$. This gives
\begin{gather}
\left|\left.\mathsf{R}_1\left(\frac{3}{2\lambda}\right)\right\vert_{n=\frac{3}{2|\lambda|}+b_0,\ m=\frac{3}{4|\lambda|}+b_1}\right|=\mathcal{O}\left(\frac{1}{|\lambda|^{-\frac{1}{2}}} e^{-\frac{3}{2|\lambda|}(1+\ln 2)}\right).
\label{Rhyper1ots1}
\end{gather}
One may therefore write (choosing $b_0$ and $b_1$ such that $\frac{3}{2|\lambda|}+b_0=[\frac{3}{2|\lambda|}]$ and $\frac{3}{4|\lambda|}+b_1=[\frac{3}{4|\lambda|}]$)
\begin{gather}
Z(0)= \frac{1}{\sqrt{\pi}}\sum_{k=0}^{[\frac{3}{2|\lambda|}]-1}\frac{(-1)^k}{k!}
\frac{\Gamma\left(\frac{1}{2}+2k\right)}{(3!)^{k}}\lambda^k \nonumber\\
\phantom{Z(0)=}{} -\frac{1}{\sqrt{2}\pi}\sum_{j=0}^{[\frac{3}{4|\lambda|}]-1}(-1)^jA_j\frac{1}{2i\pi}
\int_{d+[\frac{3}{2|\lambda|}]-i\infty}^{d+[\frac{3}{2|\lambda|}]
+i\infty}dt\left(\frac{1}{\lambda}\right)^{-t}\left(\frac{2}{3}\right)^t\frac{\pi}{\sin(\pi t)}\Gamma(t-j)\nonumber\\
\phantom{Z(0)=}{}
+\mathcal{O}\left(\frac{1}{|\lambda|^{-\frac{1}{2}}} e^{-\frac{3}{2|\lambda|}(1+\ln 2)}\right)  ,
\label{Z0hyper1ots1}
\end{gather}
which constitutes the optimally truncated hyperasymptotic expansion of $Z(0)$ at f\/irst level, in the Berry--Howls optimal truncation scheme \cite{Berry, Berry2}, that we call OTS1 in the following.

In fact, in OTS1, equation~(\ref{hyperRnm}) constitutes an asymptotic expansion which, being extracted from the exponentially suppressed remainder $\left.\mathsf{R}_0\right\vert_{n=\frac{3}{|2\lambda|}+b_0}$, allows us to conclude that at the f\/irst stage of the OTS1 hyperasymptotic approach, we extracted from the perturbative remainder a~(leading) \textit{non-perturbative asymptotic series}.

Let us stress that at f\/irst hyperasymptotic level there exists another optimal truncation scheme that leads to even better results. Indeed, instead of keeping the superasymptotic value $n=\frac{3}{2|\lambda|}+b_0$, one could have let it free: if we choose $n=\frac{a_0}{|\lambda|}+b_0$ and $m=\frac{a_1}{|\lambda|}+b_1$ (where $a_0>0$, $a_1>0$ and $b_0$ and $b_1$ are bounded), one has
\[
\left|\left.\mathsf{R}_1\left(\frac{3}{2\lambda}\right)\right\vert_{n=\frac{a_0}{|\lambda|}+b_0,\ m=\frac{a_1}{|\lambda|}+b_1}\right|
=\mathcal{O}\left(\frac{1}{|\lambda|^{-\frac{1}{2}}}\left(e^{-a_0}a_1^{a_1}
\left(\frac{3}{2}\right)^{-a_0}(a_0-a_1)^{a_0-a_1}\right)^{\frac{1}{|\lambda|}}\right),
\]
as $|\lambda|\rightarrow 0$ in $\left |\arg\left(\frac{3}{2\lambda}\right)\right |\leq\pi$.

The minimum of the remainder in this case is obtained for $a_0=3$ and $a_1=\frac{3}{2}$. This gives
\[
\left|\left.\mathsf{R}_1\left(\frac{3}{2\lambda}\right)\right\vert_{n=\frac{3}{|\lambda|}+b_0,\ m=\frac{3}{2|\lambda|}+b_1}\right|=\mathcal{O}\left(\frac{1}{|\lambda|^{-\frac{1}{2}}} e^{-\frac{3}{|\lambda|}}\right),
\]
which is more exponentially suppressed than (\ref{Rhyper1ots1}).

From this, one deduces that (choosing $b_0$ and $b_1$ such that $\frac{3}{|\lambda|}+b_0=[\frac{3}{|\lambda|}]$ and $\frac{3}{2|\lambda|}+b_1=[\frac{3}{2|\lambda|}]$)
\begin{gather}
 Z(0)=\frac{1}{\sqrt{\pi}}\sum_{k=0}^{[\frac{3}{|\lambda|}]-1}
 \frac{(-1)^k}{k!}\frac{\Gamma\left(\frac{1}{2}+2k\right)}{(3!)^{k}}\lambda^k
 -\frac{1}{\sqrt{2}\pi}\sum_{j=0}^{[\frac{3}{2|\lambda|}]-1}(-1)^jA_j\frac{1}{2i\pi}
 \int_{d+[\frac{3}{|\lambda|}]-i\infty}^{d+[\frac{3}{|\lambda|}]+i\infty}dt\left(\frac{1}{\lambda}\right)^{-t}
 \nonumber\\
 \phantom{Z(0)=}{}\times
 \left(\frac{2}{3}\right)^t\frac{\pi}{\sin(\pi t)}\Gamma(t-j)+\mathcal{O}\left(\frac{1}{|\lambda|^{-\frac{1}{2}}} e^{-\frac{3}{|\lambda|}}\right)  ,
\label{Z0hyper1ots2}
\end{gather}
which constitutes the f\/irst level of the hyperasymptotic expansion of $Z(0)$ in the Olde Daalhuis--Olver optimal truncation scheme \cite{DaalhuisOlver} that we call OTS2 in the following.

Comparing (\ref{superasymp}) and (\ref{Z0hyper1ots2}), one sees that the f\/irst level of the hyperasymptotic expansion in OTS2 therefore leads to an exponential improvement of the superasymptotic expansion. This exponential improvement is of the same order as the exponential improvement that was obtained by the superasymptotic expansion of the perturbative series. In this sense equation~(\ref{Z0hyper1ots2}) is non-perturbative.

The non-perturbative results (\ref{Z0hyper1ots1}) and (\ref{Z0hyper1ots2}) have been obtained from truncation indices $n=[\frac{a_0}{|\lambda|}]$ and $m=[\frac{a_1}{|\lambda|}]$ and we saw that the values of the coef\/f\/icients $a_0$ and $a_1$ depend on the chosen optimal truncation scheme.

It is interesting to note that although (\ref{Z0hyper1ots2}) gives a more accurate description of $Z(0)$ than~(\ref{Z0hyper1ots1}) since the remainder in OTS2 is more exponentially suppressed than the remainder in OTS1, the number of perturbative terms retained in (\ref{Z0hyper1ots2}) is at least twice the number of those retained in (\ref{Z0hyper1ots1}), which means that taking into account perturbative terms that make the perturbative series begin to diverge from the exact result leads to an improvement\footnote{This will be conf\/irmed numerically in Section~\ref{OTS}.}!

Let us f\/inish this section by stating the values obtained for both optimal truncation schemes at second hyperasymptotic level.

In OTS1, one f\/inds that the optimal values for the truncation indices are $n=[\frac{3}{2|\lambda|}]$, $m=[\frac{3}{4|\lambda|}]$ and $m'=[\frac{3}{8|\lambda|}]$, while in OTS2 these values are $n=[\frac{9}{2|\lambda|}]$, $m=[\frac{3}{|\lambda|}]$ and $m'=[\frac{3}{2|\lambda|}]$.

\subsubsection{Numerical analysis}\label{OTS}

In this section we want to compare numerical results that may be obtained from the hyperasymptotic expansion of $Z(0)$ at various hyperasymptotic levels, in both cases of optimal truncation schemes OTS1 and OTS2, with those obtained from perturbation theory in Section~\ref{pertNum}, i.e.\ for $\lambda=\frac{1}{3}$.

Recall that from (\ref{vacuum}) one gets, with a 9 decimal places precision, the ``exact'' result
\begin{gather}
\label{numeriqOTS}
Z(0)\big\vert_{\lambda=\frac{1}{3}}= 0.965560481,
\end{gather}
while the best perturbative prediction is given by (see (\ref{pertpred3}))
\begin{gather}
\label{pertpred3OTS}
Z(0)\big\vert^{\textrm{Pert}}_{\lambda=\frac{1}{3}}= 0.96555187\pm 0.00140990.
\end{gather}

{\bf First hyperasymptotic level.}
From (\ref{MBBinterpret}) (or (\ref{MBBZ0})) in OTS1 without considering the remainder integral\footnote{In all the numerical evaluations coming from hyperasymptotic formulas at f\/irst and second hyperasymptotic levels, adding the remainder integral of course always leads to the exact result (\ref{numeriqOTS}). At third level, this will not be the case due to numerical instability (see footnote \ref{DaalhuisOlver}).}, i.e.\ from  (\ref{Z0hyper1ots1}), one obtains
\begin{gather}
\label{hyper1OTS1num}
Z(0)\vert^{\textrm{OTS1}}_{\lambda=\frac{1}{3}}= 0.96552297,
\end{gather}
while in OTS2, i.e.\ from  (\ref{Z0hyper1ots2}), one has
\begin{gather}
\label{hyper1OTS2num}
Z(0)\vert^{\textrm{OTS2}}_{\lambda=\frac{1}{3}}= 0.96556291.
\end{gather}
Comparing with (\ref{pertpred3OTS}) one sees that perturbation theory gives a better prediction than (\ref{hyper1OTS1num}). This is due to the fact that the perturbative contribution to~(\ref{hyper1OTS1num}) comes from a truncation of~(\ref{asymp}) after the fourth term, while (\ref{pertpred3OTS}) is not exactly a naive truncation of the perturbative series (see~(\ref{Stieltjes})). (\ref{hyper1OTS2num}) is however closer to the exact result than the central value of~(\ref{pertpred3OTS}).

\vspace{0.3cm}

{\bf Second hyperasymptotic level.}
From (\ref{MBB4}) in OTS1 (i.e.\ with $n=[\frac{3}{2|\lambda|}]$, $m=[\frac{3}{4|\lambda|}]$ and $m'=[\frac{3}{8|\lambda|}]$) still without considering the remainder integral, one gets
\begin{gather}
\label{hyper2OTS1num}
Z(0)\vert^{\textrm{OTS1}}_{\lambda=\frac{1}{3}}= 0.96556492,
\end{gather}
while in OTS2, i.e.\ with $n=[\frac{9}{2|\lambda|}]$, $m=[\frac{3}{|\lambda|}]$ and $m'=[\frac{3}{2|\lambda|}]$,
\begin{gather}
\label{hyper2OTS2num}
Z(0)\vert^{\textrm{OTS2}}_{\lambda=\frac{1}{3}}= 0.965560477.
\end{gather}
At this hyperasymptotic level (\ref{hyper2OTS2num}) is already very close to (\ref{numeriqOTS}) while (\ref{hyper2OTS1num}) gives a prediction less good than (\ref{hyper1OTS2num}) but better than perturbation theory.

\vspace{0.3cm}

{\bf Third hyperasymptotic level.}
At this level (for which we do not give the expression of the corresponding hyperasymptotic expansion of $Z(0)$), one has no prediction from OTS1, since it would imply $m''=[\frac{3}{16|\lambda|}]=0$ (where $m''$ is the truncation index of the partial sum appearing at third hyperasymptotic level).

In OTS2, however, one has $n=[\frac{6}{|\lambda|}]$, $m=[\frac{9}{2|\lambda|}]$, $m'=[\frac{3}{|\lambda|}]$ and $m''=[\frac{3}{2|\lambda|}]$, so that
\begin{gather}
\label{hyper3OTS2num}
Z(0)\vert^{\textrm{OTS2}}_{\lambda=\frac{1}{3}}= 0.965560486,
\end{gather}
which gives the exact result with an 8 decimal places precision\footnote{\label{DaalhuisOlver}Notice that (\ref{hyper3OTS2num}) is numerically a bit less good than (\ref{hyper2OTS2num}) although it is analytically more precise. This fact is known as a numerical instability~\cite{DaalhuisOlver} and we checked that it is possible to cure it by choosing the numerically stable scheme proposed in this reference. One f\/inds in this case $Z(0)\vert_{\lambda=\frac{1}{3}}= 0.965560480$.}, although we see that $n=18$, which means that the perturbative partial sum contribution to (\ref{hyper3OTS2num}) is (see Table \ref{table1})
\begin{gather}
\label{hyper3OTS2pert}
S^{\textrm{Pert}}_{17}=-27.6964871,
\end{gather}
a value extremely far from the exact result (\ref{numeriqOTS})!

It is clear from this analysis that OTS2 is a much better optimal truncation scheme than OTS1 and that the results obtained in OTS2 give much closer value to the exact result than perturbation theory.

\section{Conclusions}

This paper was aimed to show how an interpretative procedure of divergent perturbative series, involving Mellin--Barnes representation and Borel resummation, leads to a non-perturbative asymptotic improvement of the $N$-point functions perturbative expansions, for the simple case of zero-dimensional Euclidean $\phi^4$ theory, by the appearance of hyperasymptotic expansions. We saw that these hyperasymptotic expansions are composed of interwoven partial sums whose coef\/f\/icients, in our cases of study, are linked together by a resurgence phenomenon. The non-perturbative interpretation of our results relies crucially on so-called optimal truncation schemes of these partial sums. A numerical analysis has been performed, showing the much more accurate results that one may obtain with the non-perturbative hyperasymptotic expansions, compared to the traditional perturbative approach. We also saw that on the Stokes line $\arg(\lambda)=\pi$, one may f\/ind, from hyperasymptotic expansions, imaginary contributions that are of course not obtainable from the truncated perturbative expansions.

One of the important conclusions concerns the two dif\/ferent optimal truncation schemes that were studied in this paper. Indeed, the best one (i.e.\ the one that leads to the best analytical expressions: OTS2) implies a truncation of the perturbative series that depends on the hyperasymptotic level at which the improvement is performed. This leads to the striking result that the higher the hyperasymptotic level is reached in the analysis, the more one has to take into account terms in the perturbative series. In other words, the more one wants to increase the exponential improvement, the more one also has to include perturbative contributions far in the divergent tail of the perturbative series. OTS2 then implies that, as our numerical study showed, the corresponding numerical predictions are of an amazing accuracy (see equation~(\ref{hyper3OTS2num}) compared to the exact result (\ref{numeriqOTS})), although they involve perturbative contributions that, taken independently of the non-perturbative corrections, would lead to disastrous results (see equation~(\ref{hyper3OTS2pert})). Care should however be taken concerning numerical instability issues at high hyperasymptotic level (see footnote \ref{DaalhuisOlver}).

These results clearly show that hyperasymptotic expansions are tools that could have a lot of relevance in high energy physics. In this respect, concerning the study of the Stokes phenomenon (and the smoothing of the Borel ambiguity), we would like to mention an interesting paper \cite{Pasquetti:2009jg} which appeared a few days before the end of the writing of our manuscript and dealing, among others things, with a hyperasymptotic approach of instantons in topological string theory and $c=1$ matrix models. It is however not based on Mellin--Barnes hyperasymptotic theory but rather on the hyperasymptotic approach developed by Berry and Howls~\cite{Berry2}.

\appendix

\section{General bound for the remainder\\ of an alternating asymptotic series}

Let $S_\infty=\sum\limits_{k=0}^{\infty}u_k$ be the sum of a divergent alternating asymptotic series.
In the interpretative procedure to give a meaning to this sum, one def\/ines the remainder $R_n$ as
\[
S_\infty=S_{n-1}+R_n.
\]
We therefore have
\[
R_n-R_{n+1}=u_{n},
\]
so that if $u_{n}>0$, since the asymptotic series is an alternating one, $R_{n+1}<0$ and then $R_n>0$. Therefore $R_n-R_{n+1}=|R_n|+|R_{n+1}|=|u_n|$.

On the other hand, if $u_{n}<0$, then $R_{n+1}>0$ and $R_n<0$. Therefore $R_n-R_{n+1}=-|R_n|-|R_{n+1}|=-|u_n|$.

We conclude that in all cases $|R_n|+|R_{n+1}|=|u_n|$ and that we therefore also have $|R_{n-1}|+|R_n|=|u_{n-1}|$, which means that
\[
|R_{n}|<|u_n|
\]
and
\[
|R_{n}|<|u_{n-1}|.
\]

\subsection*{Acknowledgements}

We would like to thank Santi Peris and Eduardo de Rafael for their comments, as well as one of the referees for his suggestions to improve the manuscript.
D.G.\ acknowledges f\/inancial support from CICYT-FEDER-FPA2008-01430, and the Spanish Consolider-Ingenio 2010 Program CPAN (CSD2007-00042).

\pdfbookmark[1]{References}{ref}
\LastPageEnding

\end{document}